\documentclass[12pt]{article}
\usepackage{epsfig,psfig,graphicx,rotating}
\begin{document}
\title{\bf  
STATISTICAL MECHANICS OF THE SELF-GRAVITATING GAS: I. THERMODYNAMIC
LIMIT AND PHASE DIAGRAMS}
\author{{\bf  H. J. de Vega}(a), {\bf N. S\'anchez}(b) \\ \\
(a)Laboratoire de Physique Th\'eorique et Hautes Energies, \\
Universit\'e Paris VI, Tour 16, 1er \'etage, \\ 4, Place Jussieu
75252 Paris, Cedex 05, FRANCE. \\
Laboratoire Associ\'e au CNRS UMR 7589. \\ 
(b) Observatoire de Paris,  Demirm, \\ 61, Avenue de l'Observatoire, \\
75014 Paris,  FRANCE. \\
Laboratoire Associ\'e au CNRS UA 336, \\
Observatoire de Paris et \'Ecole Normale Sup\'erieure.  }
\date{\today}
\maketitle
\begin{abstract}
We provide a complete picture to the self-gravitating non-relativistic
gas at thermal equilibrium using Monte Carlo simulations,
analytic mean field methods (MF) and low density expansions. The
system is shown to possess an infinite volume limit in the grand
canonical (GCE), canonical (CE) and microcanonical (MCE) ensembles when
$(N, V)  \to \infty$, keeping 
$N/ V^{1/3}$ fixed. We {\bf compute} the equation of state (we do not assume
it as is customary), as well as the energy, free energy, entropy, 
chemical potential, specific heats, compressibilities and speed of
sound; we analyze their properties, signs and singularities.  All
physical quantities turn out to depend on a single variable $ \eta
\equiv {G \, m^2 N \over V^{1/3} \; T}$ that is kept fixed in the $
N\to \infty $ and $ V \to 
\infty $ limit. The system is in a  gaseous phase for $ \eta < \eta_T $ and
collapses into a dense object for $ \eta > \eta_T $ in the CE with the pressure
becoming large and negative. At $ \eta \simeq \eta_T $ the isothermal
compressibility diverges. This gravitational  phase transition is
associated to the Jeans' instability. Our Monte Carlo simulations yield
$ \eta_T \simeq 1.515 . \; PV/[NT] = f(\eta) $ and all physical magnitudes 
exhibit a square root branch point at $ \eta = \eta_C >  \eta_T $.
The values of  $ \eta_T $ and $ \eta_C $ change by a few percent
with the geometry 
for large $ N $: for spherical symmetry and $ N = \infty $ (MF), we
find $ \eta_C = 1.561764\ldots $ while the Monte Carlo simulations for
cubic geometry yields $ \eta_C \simeq 1.540 $. In mean field and
spherical symmetry $c_V$ diverges as $
\left(\eta_C-\eta \right)^{-1/2} $ for $ \eta \uparrow
\eta_C$ while $c_P$ and $ \kappa_T $ diverge as $ \left(\eta_0-\eta
\right)^{-1} $ for $ \eta \uparrow \eta_0 = 1.51024\ldots  $.
The function $f(\eta)$ has a
second Riemann sheet which is only physically realized in the MCE. In the MCE, 
the collapse phase transition takes place in this second sheet near $
\eta_{MC} = 1.26 $ and the pressure and temperature are
larger in the collapsed phase than in the gaseous phase. Both
collapse phase transitions (in the CE and in the MCE) are of zeroth
order since the Gibbs free energy has a jump at the transitions. 
The MF equation of state in a sphere, $ f(\eta)$, obeys a {\bf first
order} non-linear differential equation of first kind Abel's type.  
The MF gives an extremely  accurate picture in agreement with the MC
simulations both in the CE and MCE. Since we perform the MC
simulations on a cubic geometry they describe an {\bf isothermal cube}
while the MF calculations describe an isothermal sphere.
The local properties of the gas, scaling behaviour of the particle
distribution and its fractal (Haussdorf) dimension are investigated in
the companion paper \cite{II}.
\end{abstract}
\tableofcontents
\section{Statistical Mechanics of  the Self-Gravitating  Gas}

Physical systems at thermal equilibrium are usually homogeneous. This is the 
case for gases with short range  intermolecular forces (and in absence
of external fields). In such cases the entropy is maximum when the
system homogenizes.

When long range interactions as the gravitational force are present, even
the ground state is inhomogeneous. In this case,  each element of the
substance is acted on by very strong forces due to distant
particles of the gas. Hence, regions near to and far from the boundary of the 
volume occupied by the gas will be in very different conditions, and, as a 
result, the homogeneity of the gas is destroyed \cite{llms}. The state
of maximal entropy for gravitational systems is {\bf inhomogeneous}. 
This  basic  inhomogeneity suggested us that fractal
structures can arise in a self-interacting gravitational
gas\cite{natu,prd,gal,eri,pcm}. 

The inhomogeneous character of the ground state for gravitational
systems explains why the universe is {\bf not} going towards a `thermal
death'. A `thermal death' would mean that the universe evolves towards
more and more homogeneity. This can only happen if the entropy is
maximal for an homogeneous state. Instead, it is the opposite what
happens, structures are formed in the universe through
the action of the gravitational forces as time evolves.

Usual theorems in statistical mechanics break down for inhomogeneous
ground states. For example, the specific heat may be negative in the
microcanonical ensemble (not in the canonical ensemble where it is
always positive)\cite{llms}. 

As is known, the thermodynamic limit for self-gravitating systems does
not exist in its usual form ($N\to \infty,\; V \to \infty,\; N/V = $
fixed). The system collapses into a very dense phase which is
determined by the short distance (non-gravitational) forces between
the particles.

We instead find that the thermodynamic functions exist in the {\bf dilute}
limit
\begin{equation}\label{limiT}
N\to \infty\; ,\; V \to \infty\; ,\; {N\over V^{1/3}} = \mbox{fixed}
\end{equation}
where $ V $ stands for the volume of the box containing the gas.
In such a limit, the energy $E$, the free energy and the entropy turns to be
extensive. That is, we find that they take the form of $ N $ times a
function of
$$
\eta = {G \, m^2 N \over L \; T} \quad \mbox{or} \quad
\xi = { E \, L \over G \, m^2 \, N^2}
$$
where $\eta$ and $\xi$ are  intensive variables. Namely, $\eta$ and
$\xi$ stay finite when $ N $ and $ V \equiv L^3 $ tend to infinite.  $\eta$ is
appropriate for the canonical ensemble and $\xi$ for the
microcanonical ensemble. Physical magnitudes as the specific heat,
speed of sound, chemical potential and  compressibility only depend on
$\eta$ or $\xi$. $\eta$ and $\xi$ as well as the ratio $ N/L $ are
therefore  {\bf intensive} magnitudes. The energy, the free energy, the 
Gibbs free energy and the entropy are of the form $ N $ times a
function of $\eta$. These functions of $\eta$ have a finite $ N = \infty $
limit for fixed $\eta$ (once the ideal gas contributions are
subtracted). Moreover, the dependence on $ \eta $ in all these
magnitudes express through a single universal function $ f(\eta) $. 

We study here and in the companion paper\cite{II} (called paper II in what
follows) the statistical mechanics of the self-gravitating
gas. That is,  our starting point is the partition function for
non-relativistic particles interacting through their gravitational
attraction in thermal equilibrium. We study the self-gravitating gas in the
three ensembles: microcanonical (MCE), canonical (CE) and grand
canonical (GCE). We performed calculations by three methods:

\begin{itemize}

\item{By expanding
the partition function through direct calculation in powers of $1/\xi$
and $\eta$ for the MCE and CE, respectively.  These expressions apply
in the dilute regime ($ \xi \gg 1 \, , \, \eta \ll 1 $) and become
identical for both ensembles for $ N \to \infty $. At $ \eta = 0 =
1/\xi$ we recover the ideal gas behaviour.}

\item{By performing Monte Carlo simulations both in the MCE
and in the CE.  
We found in this way that the self-gravitating gas {\bf collapses} at a
critical point which depends on the ensemble considered. As shown in
fig. 1 the collapse occurs first in the canonical ensemble (point
T). The microcanonical ensemble exhibits a larger region of stability
that ends at the point MC (fig. 1). Notice that the
physical magnitudes are identical in the common region of validity of
both ensembles within the statistical error. Beyond the critical point
T the system 
becomes suddenly extremely compact with a large negative pressure in
the CE. Beyond the point MC in the MCE the pressure and the
temperature increase suddenly and the gas collapses. 
The phase transitions at T and at MC are of zeroth order since
the Gibbs free energy has discontinuities in both cases.}

\item{By using the mean field  approach we evaluate the partition
function for large $ N $. We do this computation  in the grand canonical,
canonical and microcanonical ensembles. In the three cases the
partition function is expressed as a functional integral over a
statistical weight which depends on the (continuous) particle
density. These statistical weights are of the form of the exponential
of an `effective action' proportional to $ N $. Therefore, the $ N \to
\infty $ limit follows by the saddle point method. The saddle point is
a space dependent mean field showing the inhomogeneous character of the ground
state. Corrections to the mean field are of the order $ 1/N $ and can
be safely ignored for $ N \gg 1 $ except near the critical
points. These mean field results turned out to be in  excellent agreement
with the Monte Carlo results and with the low density expansion. }
\end{itemize}
We calculate the saddle point (mean field)  for spherical symmetry
and we obtain from it the various physical
magnitudes (pressure, energy, entropy, free energy, specific heats,
compressibilities, speed of sound and particle density).
Furthermore, we compute the determinants of small fluctuations 
around the saddle point solution for spherical symmetry for the three
statistical ensembles in paper II. 

When any small fluctuation around the saddle point decreases the
statistical weight in the functional integral, the saddle point is
dominating the integral and the mean field approach is fully valid. 
In that case the  determinant of small fluctuations is positive. A
negative determinant of small fluctuations indicates that some
fluctuations around the saddle point are increasing the statistical
weight in the functional integral and hence the saddle point {\bf does
not} dominate the partition function. The mean field approach cannot be
used when the determinant of small fluctuations is negative. 

The zeroes of the small fluctuations determinant determine the  position of the
critical points for the three statistical ensembles. 
The Monte Carlo simulations for the CE and the MCE show that the
self-gravitating gas collapses near the critical points obtained from
mean field. 

The saddle point solution is identical for the three statistical
ensembles. This is not the case for the fluctuations around it. The
presence of constraints in the CE (on the number of particles) and in
the MCE (on the energy and the number of particles) changes the
functional integral over the quadratic fluctuations with respect to
the GCE. 

The saddle point of the partition function turns out to coincide with the
hydrostatic treatment of the self-gravitating
gas\cite{chandra}-\cite{bt}. (Usually known as the `isothermal sphere'
in the spherically symmetric case).

Our Monte Carlo simulations are performed in a cubic geometry. The
equilibrium configurations obtained in this manner can thus be called
the {\bf `isothermal cube'}.

\medskip

We find for spherical symmetry: $ \eta^R_{GC} = 0.797375\ldots , \;
\eta^R_{C} = 2.517551\ldots $ and $ \eta^R_{MC} = 2.03085\ldots $.
The variable $ \eta^R $ appropriate for a spherical
symmetry is defined as $ \eta^R \equiv {G \, m^2 N \over R \; T}
= \eta \; \left({4\pi\over 3}\right)^{1/3} =  1.61199\ldots \; \eta$.

\medskip

The values  of $ \eta_T $ and  $ \eta_C $ change by a few 
percent with the geometry and with the number of particles (for large
$ N > 500 $). For spherical symmetry and $ N = \infty $ (mean field)
we obtain  $ \eta_C = \left({3 \over 4\pi }\right)^{1/3} \eta^R_{C} =
1.56176\ldots $. Our Monte Carlo simulations 
yield $ \eta_T \simeq 1.515$. We find from the mean field approach
that the isothermal compressibility diverges at $ \eta= \eta_0  =
1.51024\ldots \simeq \eta_T $ for spherical symmetry.  

\medskip

The conclusion being that  the MF correctly describes the
thermodynamic limit except near the critical points (where the small
fluctuations determinant vanishes);
the MF is valid for $N|\eta-\eta_{crit}|\gg 1$. The vicinity of the
critical point should be studied in a double scaling limit $N \to
\infty,\; \eta \to \eta_{crit}$. 

\bigskip

In summary, the picture we get from 
our calculations using these three  methods show that the self-gravitating gas
behaves as a perfect gas for  $ \eta \to 0, \; 1/\xi\to 0 $. When $
\eta $ and $ 1/\xi $ grow, the gas becomes denser till it suddenly
condenses into a high density object at a critical point GC, C or MC
depending upon the statistical ensemble chosen. In the Monte Carlo
simulations for the CE the collapse takes place at the point T
slightly before $ \eta_C $. $ \eta $ is related with the
Jeans' length $ d_J $ of the gas through $ \eta = 3 ( L/d_J )^2
$. Hence, when $ \eta $ goes beyond $ \eta_T $, the length of the
system becomes larger than $ d_J / \sqrt{\eta_T /3} $. The collapse at
T in the CE is therefore a manifestation of the Jeans' instability. 
The saddle point ceases to describe the physics at C since the
determinant of fluctuations for the CE vanishes there. 

\bigskip

In the MCE, the determinant of fluctuations vanishes at the point MC.
The physical states
beyond MC are collapsed configurations shown by the Monte Carlo
simulations [see fig. \ref{colmc}]. Actually, the gas collapses in the
Monte Carlo simulations slightly before the mean field prediction for
the point MC. The phase transition at the critical point MC is
the so called gravothermal catastrophe \cite{lynbell2}.

\bigskip

The gravitational interaction being attractive without lower
bound, a short distance cut-off ($ A $) must be introduced in order
to give a meaning to the partition function. We take the gravitational
force between particles as $ -G \; m^2 /r^2 $ for $ r > A $ and zero
for $ r< A $ 
where $ r $ is the distance between the two particles. We show that
the cut-off effects are negligible in the $ N = \infty $ limit. That
is, {\bf once} we set $ N = \infty $ with fixed $ \eta $,
{\bf all} physical quantities are {\bf finite} in the
zero cut-off limit ($ A = 0$). The cut-off effects are of the order $
A^2/L^2 $ and can be safely ignored.

\bigskip

All physical quantities are expressed in terms of  $f(\eta)$. Besides
computing  $f(\eta)$ numerically in the mean field approach, we obtain
analytic results about it from the Abel's equation. There is a square
root branch point in $f(\eta)$ at $ \eta_C $. The specific heat is
positive in the first sheet and negative in the second sheet. This
second sheet is only  physically realized in the microcanonical
ensemble (MCE). [The 
specific heat is positive definite in the CE]. $f(\eta)$ has infinitely
many branches in the $\eta$ plane but only the first two are
physically realized. Beyond MC the states described by the mean field
saddle point are
unstable. We plot and analyze the equation of state, the
energy, the entropy, the free energy, $ c_V $, $ c_P $, the isothermal
compressibility and the speed of sound [figs. \ref{fig8}-\ref{fig3}].
Most of these physical magnitudes were not previously
computed in the literature as functions of $ \eta $.

We find analytically the behaviour of $f(\eta)$ near $ \eta_C $ in
mean field,
$$
f_{MF}(\eta^R) \buildrel{ \eta^R \uparrow \eta^R_C}\over= \frac13 +
0.213738\ldots \sqrt{\eta^R_C-\eta^R} +0.172225\ldots\; (\eta^R_C-\eta^R)
+ {\cal O}\left[(\eta^R_C-\eta^R)^{3/2}\right] \; .
$$
This shows that the specific heat at constant volume diverges as $
\left(\eta_C-\eta \right)^{-1/2} $ for $ \eta^R \uparrow
\eta^R_C$. The specific heat at constant pressure and the isothermal
compressibility diverge at $ \eta_0 $ as $ \left(\eta_0-\eta
\right)^{-1} $. These mean field results apply for $ |\eta-\eta_C| \ll
1 \ll N|\eta-\eta_C| $. Fluctuations around mean field can be
neglected in such a regime.

The Monte Carlo calculations permit us to obtain  $f(\eta)$ in the
collapsed phase. Such result (which is cutoff dependent) cannot be
obtained in the mean field approach. The mean field only provides
information (as $f(\eta)$) in the gas phase. 

For the self-gravitating gas, we find that the Gibbs free energy $ \Phi
$ {\bf is not} equal to $N$ times the chemical potential and that the
thermodynamic potential $ \Omega $ {\bf is not} equal to $ - PV $ as
usual\cite{llms}. This is a consequence of the dilute thermodynamic
limit $ N \to \infty, \; L\to \infty, \; N/L=$fixed.  

\bigskip

We compute {\bf local} properties of the gas in paper II. That is, the
local energy density $ \epsilon(r) $, local particle density and local
pressure. Furthermore, we  analyze the scaling behaviour of the particle
distribution and its fractal (Haussdorf) dimension.

\bigskip

This paper is organized as follows. In section II
we present the statistical mechanics of the self-gravitating gas in
the microcanonical ensemble, in sec. III we do the analogous
presentation for the canonical ensemble and in sec. IV we contrast the
results for the CE and the MCE.  Sec. V contains the results
from Monte Carlo simulations and we develop in sec. VI the mean
field approach. In sec. VII we present the results for intensive
magnitudes. Discussion and remarks are presented in section VIII whereas
appendixes A-C contain relevant mathematical developments. 

\section{Statistical Mechanics  of the  Self-Gravitating Gas: the
microcanonical ensemble}

We investigate in this section  an isolated  set of   $N$ 
non-relativistic particles with mass $m$ interacting 
 through Newtonian gravity with total energy $ E $. That is, a
self-gravitating gas in the microcanonical  ensemble. We  assume the
system being  on a cubic box of side $ L $ just for simplicity. We
consider spherical symmetry in sec. VI. Please notice that we {\bf
never} use periodic boundary conditions.

At short distances,  the particle interaction for the self-gravitating
gas in physical situations is not gravitational. Its
exact nature depends on the problem under consideration (opacity limit,
Van der Waals forces for molecules etc.). 
We shall just assume a repulsive short distance potential, that is,
\begin{equation}\label{defva}
v_A(|{\vec q}_l - {\vec q}_j|) = - {1 \over |{\vec q}_l - {\vec q}_j|_A } =
\left\{ \begin{array}{c}  - {1 \over |{\vec q}_l - {\vec q}_j|} \quad
\mbox{for} \; |{\vec q}_l - {\vec q}_j| \ge A \cr 
 \cr +{1 \over A} \quad \mbox{for}\; |{\vec q}_l - {\vec q}_j| \le A
\end{array} \right. 
\end{equation}
where $ A << L $ is the short distance cut-off. 

The presence of the repulsive short-distance interaction prevents the
collapse (here  unphysical) of the self-gravitating gas. In the situations we
are interested to describe (interstellar medium, galaxy distributions)
the collapse situation is unphysical. 

The entropy  of the system can be written as
\begin{equation}\label{smc}
S(E,N)  = \log\left\{ {1 \over N !}\int\ldots \int
\prod_{l=1}^N{{d^3p_l\, d^3q_l}\over{(2\pi)^3}}\; 
\delta\!\left[E - \sum_{l=1}^N\;{{p_l^2}\over{2m}} - 
U({\vec q}_1, \ldots {\vec q}_N) \right] \right\} 
\end{equation}
where
\begin{equation}\label{uq}
U({\vec q}_1, \ldots {\vec q}_N) = - G \, m^2 \sum_{1\leq l < j\leq N} 
{1 \over { |{\vec q}_l - {\vec q}_j|_A}}
\end{equation}
and $G$ is Newton's gravitational constant.

In order to compute the integrals over the momenta $p_l, \; (1 \leq l
\leq N) $, we introduce the variables,
$$
{\vec \rho}_i = {1 \over \sqrt{2m}} \, {\vec p}_i \; .
$$
We can now integrate over the angles in $ 3 N $ dimensions,
\begin{eqnarray}
&&\int_{-\infty}^{+\infty}\ldots \int_{-\infty}^{+\infty}\prod_{l=1}^N\;{
d^3p_l\over{(2\pi)^3}} \;  
\delta\!\left[E - \sum_{l=1}^N\;{\vec\rho}_l^{\, 2} - U({\vec q}_1, \ldots
{\vec q}_N) \right] \cr \cr
&=& \left({\sqrt{2m} \over 2\pi}\right)^{3N}\; {2 \pi^{3N/2} \over
\Gamma\left( {3N \over 2} \right) }\; \int_0^{\infty} \rho^{3N-1}\,
d\rho \; \delta\!\left[E - \rho^2 - U({\vec q}_1, \ldots {\vec q}_N)\right]
\cr \cr
&=& \left( {m \over 2\pi}\right)^{3N/2}{ 1 \over \Gamma\left( {3N
\over 2} \right) } \; \left[E - U({\vec q}_1, \ldots {\vec q}_N)
\right]^{3N/2 -1}  \; \theta\left[E - U({\vec q}_1, \ldots {\vec q}_N) \right] 
\end{eqnarray}
The delta function in the energy thus becomes the constraint of a
positive kinetic energy $ E - U({\vec q}_1, \ldots {\vec q}_N) > 0 $.
We then get for the entropy,
\begin{equation}\label{sq}
S(E,N)  = \log\left\{ {\left({ m \over 2\pi}\right)^{3N/2} \over N !\,
\Gamma\left( {3N \over 2} \right)}\, 
\int_0^L\ldots \int_0^L \prod_{l=1}^N\; d^3q_l \; 
\left[E - U({\vec q}_1, \ldots {\vec q}_N) \right]^{3N/2
-1}\theta\left[E - U({\vec q}_1, \ldots {\vec q}_N ) \right] \right\} 
\end{equation}

It is convenient to introduce the dimensionless variables $ {\vec r}_l
,\;  1\leq l \leq N $ making explicit the volume dependence  as
\begin{eqnarray}\label{variar}
{\vec q}_l &=& L \; {\vec r}_l \quad , \quad {\vec r}_l =(x_l,y_l,z_l)
\;, \cr \cr
0&\leq& x_l,y_l,z_l \leq 1\; .
\end{eqnarray}
That is, in the new coordinates the gas is inside a cube of unit volume.

The entropy then becomes
\begin{eqnarray}\label{sN}
S(E,N)  &=& \log\left\{ {N^{3N-2} \, m^{9N/2-2} \, L^{3N/2 +1} \, G^{3N/2 -1}
\over N !\, \Gamma\left( {3N \over 2} \right)\, {(2\pi)}^{3N/2}}\right.\\ \cr
&&\left.
\int_0^1\ldots \int_0^1 \prod_{l=1}^N\; d^3r_l \; 
\left[\xi + {1 \over N}u({\vec r}_1, \ldots, {\vec r}_N) \right]^{3N/2
-1}\theta\left[\xi + 
{1 \over N}u({\vec r}_1, \ldots, {\vec r}_N) \right] \right\} \nonumber
\end{eqnarray}
where we introduced the  dimensionless variable $\xi$,
\begin{equation}\label{tzi}
\xi \equiv { E \, L \over G \, m^2 \, N^2}
\end{equation}
and
\begin{equation}\label{defu}
u({\vec r}_1, \ldots, {\vec r}_N) \equiv {1 \over
N}\sum_{1\leq l < j\leq N} {1 \over { |{\vec r}_l - {\vec r}_j|_a}}\; .
\end{equation}
where $ a \equiv A/L \ll 1 $. 

Let us define the coordinate partition function in the microcanonical
ensemble as
\begin{equation}\label{defw}
w(\xi,N)\equiv \int_0^1\ldots \int_0^1 \prod_{l=1}^N\; d^3r_l \; 
\left[\xi + {1 \over N}u({\vec r}_1, \ldots, {\vec r}_N) \right]^{3N/2
-1}\theta\left[\xi + 
{1 \over N}u({\vec r}_1, \ldots, {\vec r}_N) \right]\; .
\end{equation}
Therefore,
$$
S(E,N)  = \log\left[ {N^{3N-2} \, m^{9N/2-2} \, L^{3N/2 +1} \, G^{3N/2 -1}
\over N !\, \Gamma\left( {3N \over 2} \right)\, {(2\pi)}^{3N/2}}\right]
+ \log w(\xi,N) \; .
$$
We can now compute the thermodynamic quantities, temperature and
pressure through the standard thermodynamic relations
\begin{equation}\label{termo}
{1 \over T} = \left( {\partial S \over \partial E} \right)_V \quad
\mbox{and} \quad p = T \left( {\partial S \over \partial V} \right)_E
\; ,
\end{equation}
where $ V \equiv L^3 $ stands for the volume of the system and $ p $
is the external pressure on the system.

We obtain the temperature as a function of $ E $ and $ \xi $
from eqs.(\ref{sN}) and (\ref{termo}) 
\begin{equation}\label{1sT}
{1 \over T} = { \xi \over E} {\partial \over \partial \xi}\log  
w(\xi,N) = { 3 N \xi \over 2 E } \left[ 1 - {2 \over 3N} \right]
< {1 \over \xi + {1 \over N}u(.) }>\; .
\end{equation}
where
\begin{equation}\label{valm}
< {1 \over \xi + {1 \over N}u(.) }> \equiv { \int_0^1\ldots \int_0^1
\prod_{l=1}^N\; d^3r_l \;  
\left[\xi + {1 \over N}u({\vec r}_1, \ldots, {\vec r}_N) \right]^{3N/2
-2}\theta\left[\xi + 
{1 \over N}u({\vec r}_1, \ldots, {\vec r}_N) \right] \over
\int_0^1\ldots \int_0^1 \prod_{l=1}^N\; d^3r_l \; 
\left[\xi + {1 \over N}u({\vec r}_1, \ldots, {\vec r}_N) \right]^{3N/2
-1}\theta\left[\xi + 
{1 \over N}u({\vec r}_1, \ldots, {\vec r}_N) \right]}
\end{equation}

The equation of state follows from eqs.(\ref{sN}) and (\ref{termo}) 
\begin{equation}\label{eqes}
{pV \over N T} = \frac12 + \frac1{3N} +  { \xi \over 3\,N} {\partial
\over \partial \xi}\log w(\xi,N) = \frac12 \left( 1 + {2 \over 3 N} \right) +
{\xi \over 2 } < {1 \over \xi + {1 \over N}u(.) }>\left[ 1 - {2 \over
3N} \right] \; .
\end{equation}

We are interested in the large size limit where $ N \to \infty, \; L
\to \infty $ and $ E \to \infty $. We consider that $ \xi =  { E \, L
\over G \, m^2 \, N^2} $ stays fixed in such limit. That is, we assume
$ E/N $ and $ L/N $ bounded and nonzero when $ E,\; L $ and $ N \to
\infty $. We shall see below that such limit is meaningful.

It is possible to write the energy and the equation of state in terms
of a single function
\begin{equation}\label{gzi}
g(\xi) \equiv { \xi \over N} {\partial \over \partial \xi}\log  
w(\xi,N) = { 3 \xi \over 2} < {1 \over \xi + {1 \over N}u(.) }>
\left[ 1 - {2 \over 3N} \right] \; .
\end{equation}

We find from eqs.(\ref{1sT}), (\ref{eqes}) and (\ref{gzi}),
\begin{eqnarray}\label{EypV}
{pV \over N T} &=& \frac12 + \frac13 g(\xi) + \frac1{3N} \cr \cr
{ E \over NT}  &=&  g(\xi) \; .
\end{eqnarray}
We obtain the virial theorem by eliminating $ g(\xi) $ in the
eqs.(\ref{EypV}) 
\begin{equation}\label{virial}
pV = {N T \over 2} + {E + T \over 3}\; .
\end{equation}
where the term $ T/3 $ can be neglected for large $ N $.

In the case of a perfect gas (no gravity) we have $ u(.) \equiv 0 , \; g(\xi) =
\frac32 , \; pV = NT $ and $ E = \frac32 NT $ as it must be. 

The function $  g(\xi) $ is computed by
Monte Carlo simulations, mean field methods and,  in the weak field
limit $ \xi << 1 $, is calculated analytically in powers of $
1/\xi $ in subsection II.A.

\bigskip

The specific heat per particle is given by
$$
c_V = {T \over N} \left( {\partial S \over \partial T} \right)_V =
\frac1{N  \left( {\partial T \over \partial E} \right)_V   } 
$$
Hence, using eq.(\ref{EypV}) yields
$$
c_V = {  g(\xi) \over 1 - { \xi  g'(\xi) \over  g(\xi) }}\quad
\mbox{or}\quad {1 \over c_V} = {d \over d\xi}\left[ {\xi \over
g(\xi)}\right]
$$

We can relate the specific heat $ c_V $ with the fluctuations as
follows. We can express $   g(\xi) $ as an average value using
eqs.(\ref{valm}) and (\ref{gzi})
$$
{ \xi \over  g(\xi)} = {2 \over 3 - \frac2{N}}
{ \int_0^1\ldots \int_0^1 \prod_{l=1}^N\; d^3r_l \; 
\left[\xi + {1 \over N}u({\vec r}_1, \ldots, {\vec r}_N) \right]^{3N/2
-1}\theta\left[\xi + {1 \over N}u({\vec r}_1, \ldots, {\vec r}_N) \right] \over
\int_0^1\ldots \int_0^1 \prod_{l=1}^N\; d^3r_l \;
\left[\xi + {1 \over N}u({\vec r}_1, \ldots, {\vec r}_N) \right]^{3N/2
-2}\theta\left[\xi + {1 \over N}u({\vec r}_1, \ldots, {\vec r}_N) \right]}
$$
Computing the derivative with respect to  $\xi$ yields,
\begin{equation}
{1 \over c_V}=
\frac23 - {\left(\Delta {1 \over \xi + {1 \over N}u(.) }\right)^2
\over <{1 \over 
\xi + {1 \over N}u(.) } >^2} + {\cal O}\left({1 \over N}\right)
\end{equation}
where 
$$
\left(\Delta {1 \over \xi + {1 \over N}u(.) }\right)^2 \equiv N\left\{
<{1 \over 
[ \xi + {1 \over N}u(.) ]^2 } >- <{1 \over \xi + {1 \over N}u(.) } >^2\right\}
$$
is of order $ N^0 $ for $ N \to \infty $.
[Notice that in the calculation of the fluctuations we must keep the $ 1/N $
corrections till the end]. 

We can express $ c_V $ in terms of the fluctuations of the inverse
temperature $ \beta \equiv 1/T $ using eq.(\ref{1sT}):
\begin{equation}
{1 \over c_V}= \frac23 - \left({\Delta \beta   \over \beta}\right)^2
\end{equation}

It must be noticed that in the microcanonical ensemble, $ c_V $ may be
positive as well as negative. In fact, it becomes negative
when the fluctuations are large enough [see sec. V and VI]. 

We see that extensivity holds here in an specific way. $ T, \; S/N $
and $ pV/N $ are of order one for $ N \to \infty $ {\bf provided} $
\xi $ stays fixed in such limit. That is, we must keep $ E/N $ and $
L/N $ fixed in the $ N \to \infty $ limit.

\subsection{The diluted regime: $ \xi >> 1$}

We can obtain the thermodynamic quantities as a series  in powers
of $ 1/\xi $ just expanding the integrand in eq.(\ref{defw}).

We find
\begin{eqnarray}\label{w2o}
w(\xi,N)&{\buildrel{ \xi \to \infty}\over =} &\xi^{3N/2-1} \left\{1 
 + {9 \, b_0 \, N \over 2 \xi}\left(1 - { 2 \over
3N}\right)\left(1 - { 1 \over N}\right) \right. \cr \cr
&+&\left. {9 \over 8 \, \xi^2}\left(1 - { 2 \over3N}\right)
\left(1 - { 4 \over3N}\right)\left[9 \,N^2 \, b_0^2 \left(1 - { 1
\over N}\right) \left(1 - { 2 \over N}\right)\left(1 - { 3 \over
N}\right) \right. \right. \cr \cr
 &+&\left.\left.  N\, b_1 \left(1 - { 1 \over N}\right) \left(1 - { 2
\over N}\right) + \frac12 \, b_2 \left(1 - { 1 \over N}\right) \right]
+{\cal O}(\xi^{-3}) \right\}
\end{eqnarray}
where $ b_0, \; b_1 $ and $b_2 $ are pure numbers,
\begin{eqnarray}\label{b1b2}
b_0 &=& \frac16 \int_0^1 \int_0^1 { {d^3 r_1 \; d^3 r_2} \over
{ |{\vec r}_1 - {\vec r}_2|}}\cr \cr
b_1 &=& \int_0^1 \int_0^1 { {d^3 r_1 \; d^3 r_2\; d^3 r_3} \over
{ |{\vec r}_1 - {\vec r}_2| |{\vec r}_1 - {\vec r}_3|}}\cr \cr
b_2 &=&\int_0^1 \int_0^1 { {d^3 r_1 \; d^3 r_2} \over
{ |{\vec r}_1 - {\vec r}_2|^2}}
\end{eqnarray}

For the cubic geometry chosen, it takes the value
$$
b_0^{cube} = \frac43 \;  \int_0^1 (1-x) \, dx \int_0^1 (1-y)\, dy\int_0^1 
{(1-z)\, dz \over \sqrt{x^2 + y^2 + z^2}} = 0.31372\ldots\; .
$$

For a sphere of unit volume we find
\begin{eqnarray}\label{esfera}
b_0^{sphere} &=& \frac15 \, \left({4\pi\over 3}\right)^{1/3} = 0.32239839\ldots
\quad , \cr \cr
b_1^{sphere} &=& \frac{51}{35}\, \left({4\pi\over 3}\right)^{2/3}=
3.786412026\ldots \; ,\cr \cr
b_2^{sphere} &=& \frac94\, \left({4\pi\over 3}\right)^{2/3} =
5.846665629\ldots \; .
\end{eqnarray}

We see that the coefficient $ b_0 $ for cubic and spherical geometries only
differ by about $3\%$.

We thus find from eq.(\ref{w2o}) in the $ N \to \infty $ limit
\begin{equation}\label{limw}
\mbox{lim}_{N\to \infty}\frac{1}{N} \log w(\xi,N) = \frac32 \log\xi+
{ 9\, b_0\over 2 \xi} +  {9 \over 8\, \xi^2}\left(  b_1  -
42 b_0^2 \right) + {\cal O}(\xi^{-3})
\end{equation}

\bigskip

We considered  here these integrals in the zero cutoff limit since $ b_0,
\; b_1 $ and $ b_2 $ have finite zero cutoff limits. It is easy to see
that their finite cutoff values behave as 
\begin{equation}\label{bcut}
b_0(a) - b_0 = {\cal O}(a^2) \quad , \quad
b_1(a) - b_1 = {\cal O}(a^4) \quad , \quad
b_2(a) - b_2 = {\cal O}(a) 
\end{equation}

\bigskip

Inserting eq.(\ref{limw}) into eq.(\ref{gzi}) yields,
\begin{equation}
 g(\xi) = \frac32 -{ 9\, b_0\over 2 \xi} - {9 \over 4\, \xi^2}\left(b_1  - 
42 \, b_0^2 \right) + {\cal O}(\xi^{-3})
\end{equation}
and
\begin{equation}\label{microdilu}
{pV \over N T}= 1  -{ 3\, b_0\over 2 \xi} - {3 \over 4\, \xi^2}\left(b_1  - 
42 \, b_0^2 \right) + {\cal O}(\xi^{-3}) \; .
\end{equation}
We see that after letting $ N \to \infty $ the zero cutoff limit is
finite. We further discuss this important issue in the next section.

\section{Statistical Mechanics  of the  Self-Gravitating Gas: the canonical
ensemble}

We investigate in this section  the  self-gravitating gas considered
in the previous section but in thermal
equilibrium at temperature $ T \equiv \beta^{-1} $. That is, we work
in the canonical ensemble where the
system of $ N $ particles is not isolated but in contact with a thermal bath at
temperature $ T $. We keep assuming the gas being  on a
cubic box of side $ L $.

The partition function of the system can be written as

\begin{equation}\label{fp}
{\cal Z}_C(N,T) = {1 \over N !}\int\ldots \int
\prod_{l=1}^N\;{{d^3p_l\, d^3q_l}\over{(2\pi)^3}}\; e^{- \beta H_N}
\end{equation}
where
\begin{equation}\label{hamic}
H_N = \sum_{l=1}^N\;{{p_l^2}\over{2m}} - G \, m^2 \sum_{1\leq l < j\leq N}
{1 \over { |{\vec q}_l - {\vec q}_j|_A}}
\end{equation}
$G$ is Newton's gravitational constant.

Computing the integrals over the momenta $p_l, \; (1 \leq l \leq N) $

$$
\int_{-\infty}^{+\infty}\;{{d^3p}\over{(2\pi)^3}}\; e^{- {{\beta
p^2}\over{2m}}} = \left({m \over{2\pi \beta}}\right)^{3/2}
$$

yields

\begin{equation}\label{gfpc}
\displaystyle{
{\cal Z}_C(N,T) = {1 \over N !} \left({m \over{2\pi \beta}}\right)^{\frac{3N}2}
\; \int_0^L\ldots \int_0^L
\prod_{l=1}^N d^3q_l\;\; e^{ \beta G \, m^2 \sum_{1\leq l < j\leq N}
{1 \over { |{\vec q}_l - {\vec q}_j|_A}} }}\; .
\end{equation}
We make now explicit the volume dependence introducing the 
variables $ {\vec r}_l ,\;  1\leq l \leq N $ defined in eq.(\ref{variar}).
The partition function takes then the form,
\begin{equation}\label{fp2}
{\cal Z}_C(N,T) = {1 \over N !}\left({m T L^2\over{2\pi}}\right)^{\frac{3N}2}
\; \int_0^1\ldots \int_0^1
\prod_{l=1}^N d^3r_l\;\; e^{ \eta \; u({\vec r}_1,\ldots,{\vec r}_N)}\; ,
\end{equation}
where we introduced the  dimensionless variable $ \eta $ 
\begin{eqnarray}\label{defeta}
\eta &\equiv& {G \, m^2 N \over L \; T}
\end{eqnarray}
and $ u( {\vec r}_1, \ldots,{\vec r}_N) $ is defined by eq.(\ref{defu}).
Recall that
\begin{equation}\label{Upot}
 U \equiv - {G \, m^2 N \over L}\; u({\vec r}_1,\ldots,{\vec r}_N) \; ,
\end{equation}
is the potential energy of the gas.

The free energy  takes then the  form,
\begin{equation}\label{flib}
F = -T \log {\cal Z}_C(N,T) = -N T \log\left[ { e V \over N} \left({mT\over
2\pi}\right)^{3/2}\right]  - T \; \Phi_N(\eta) \; ,
\end{equation}
where
\begin{equation}\label{FiN}
\Phi_N(\eta) = \log \int_0^1\ldots \int_0^1
\prod_{l=1}^N d^3r_l\;\; e^{ \eta \; u({\vec r}_1,\ldots,{\vec r}_N)}\; ,
\end{equation}
The derivative of the function $ \Phi_N(\eta) $ will  be computed by
Monte Carlo simulations, mean field methods and,  in the weak field
limit $ \eta << 1 $, it will be calculated analytically.

We get for the pressure of the gas,
\begin{equation}\label{pres}
p = - \left({ \partial F \over  \partial V}\right)_T = {N T \over V} -
{\eta \, T \over 3 \, V} \; \Phi_N'(\eta)\; .  
\end{equation}
[Here, $ V \equiv L^3 $ stands for the volume of the box and $ p $
is the external pressure on the system.].
We see from eq.(\ref{FiN}) that $ \Phi_N(\eta) $ increases with $ \eta
$ since $ u(.) $ is positive. Therefore, the second term in
eq.(\ref{pres}) is a {\bf negative} correction to the perfect gas
pressure $ {N T \over V} $.  

The mean value of the potential energy $ U $ can be written from
eq.(\ref{Upot}) as
\begin{equation}\label{umed}
<U> = - T \eta  \; \Phi_N'(\eta)
\end{equation}
Combining eqs.(\ref{pres}) and (\ref{umed}) yields the virial theorem,
\begin{equation}\label{virial2}
{p V \over N T} = 1 +{ <U>\over 3 N T}\quad \mbox{or} \quad
{p V \over N T} = \frac12 + { E \over 3 N T} \; ,
\end{equation}
where we use that the average value of the kinetic energy of the gas
is $ \frac32 NT $.

A more explicit form of the equation of state is
\begin{equation} \label{estado}
{p V \over N T} = 1- {\eta \over 3 N} \; \Phi_N'(\eta)\; ,
\end{equation}
where
\begin{eqnarray}\label{fiprima}
\Phi_N'(\eta) &=& e^{-\Phi_N(\eta)} \; \int_0^1\ldots \int_0^1
\prod_{l=1}^N d^3r_l\;  u({\vec r}_1,\ldots,{\vec r}_N)\; e^{ \eta
u({\vec r}_1,\ldots,{\vec r}_N)}\cr 
\cr  &=& \frac12(N-1)  \; e^{-\Phi_N(\eta)} \; \int_0^1\ldots \int_0^1
\prod_{l=1}^N d^3r_l\;{1 \over |{\vec r}_1-{\vec r}_2|_a}\; e^{ \eta
u({\vec r}_1,\ldots,{\vec r}_N)} \; .
\end{eqnarray}
This formula indicates that $ \Phi_N'(\eta) $ is of order $ N $ for
large $ N $. Monte Carlo simulations  as well as analytic calculations
for small $ \eta $  show that this is indeed the case. In conclusion,
we can write the equation of state of the self-gravitating gas as
\begin{equation}\label{pVnT}
{p V \over N T} = f(\eta) \quad ,     
\end{equation}
where the function 
$$ 
f(\eta) \equiv  1- {\eta \over 3 N} \; \Phi_N'(\eta)\; ,
$$ 
is {\bf independent} of $ N $  for large $N$ and fixed $ \eta $.
[In practice, Monte Carlo simulations show that $ f(\eta) $ is
independent of $ N $ for  $ N > 100 $]. 

We get in addition,
\begin{equation}\label{Ueta}
<U>= -3 N T\;[ 1-  f(\eta)]\;.
\end{equation}
In the dilute  limit, $ \eta \to 0 $ and we find the perfect gas
value
$$
f(0) = 1\; .
$$
Equating eqs.(\ref{estado}) and (\ref{pVnT}) yields,
$$
\Phi_N(\eta)= 3N \, \int_0^{\eta} dx \, { 1 - f(x) \over x}\; .
$$
Relevant thermodynamic quantities can be expressed in terms of the
function $ f(\eta) $. We find for the free energy from
eq.(\ref{flib}),
\begin{equation}\label{enlib}
F =  F_0- 3NT\; \int_0^{\eta} dx \; { 1 - f(x) \over x}\; . 
\end{equation}
where 
\begin{equation}\label{Fcero}
F_0 = -N T \log\left[{ e V \over N} \left({mT\over
2\pi}\right)^{3/2}\right]
\end{equation}
is the free energy for an ideal gas.

We find for the total energy $ E $, chemical potential $ \mu $ and
entropy $ S $ the following expressions,
\begin{eqnarray}\label{ecan}
E &=&  3NT\left[  f(\eta) -\frac12\right] \; , \\  \cr
\label{poqui}
\mu &=& \left({ \partial F \over  \partial N}\right)_{T,V} = 
-T \log\left[{V \over N}\left({mT\over 2\pi}\right)^{3/2}\right]
- 3T[1 - f(\eta)] - 3T\; \int_0^{\eta} dx \, { 1 - f(x) \over x} 
\; , \cr \cr
\label{entro}
S &=& - \left({ \partial F \over  \partial T}\right)_V \cr \cr
&=& S_0  + 3N\left[
\int_0^{\eta} dx \, { 1 - f(x) \over x}+  f(\eta) -1 \right] \; ,
\end{eqnarray}
where
$$
S_0 = -{F_0 \over T} +\frac32 N \; . 
$$
is the entropy of the ideal gas.

Notice that here the Gibbs free energy 
\begin{equation}\label{gibbs}
\Phi = F + pV = F_0 + NT \left[ f(\eta) - 3 \,  \int_0^{\eta} dx \, {
1 - f(x) \over x}  \right] \; ,
\end{equation}
is {\bf not} proportional to the chemical potential. That is,
here $ \Phi \neq \mu \, N $ and we have instead,
\begin{equation}\label{gibbs2}
\Phi - \mu \, N = 2 NT \left[ 1-f(\eta) \right] \; .
\end{equation}
This relationship differs from the customary one (see \cite{llms}) due
to the fact that the dilute scaling relation $ N \sim L $ holds here
instead of the usual one $ N \sim L^3 $. The usual relationship is
only recovered in the ideal gas limit $ \eta = 0 $.  

\bigskip

The specific heat at constant volume takes the form\cite{llms},
\begin{eqnarray}\label{ceV}
c_V &=& {T \over N}  \left({ \partial S \over  \partial T}\right)_V
= 3 \left[  f(\eta)-\eta \; f'(\eta) -\frac12 \right]\; .
\end{eqnarray}
where we used eq.(\ref{entro}).
This quantity is also related to the fluctuations of the potential
energy $(\Delta U)^2$ and it is positive defined in the canonical ensemble,
\begin{equation}\label{cV}
c_V =\frac32 + (\Delta U)^2 \; .
\end{equation}
Here,
\begin{equation}\label{delu}
(\Delta U)^2 \equiv {{<U^2>-<U>^2}\over N \; T^2}= 3 \left[
f(\eta)-\eta \; f'(\eta) - 1 \right]\; .
\end{equation}

The specific heat at constant pressure is given by \cite{llms}
\begin{equation}\label{cpcv}
c_P = c_V -  {T \over N}{{ \left({ \partial p \over  \partial
T}\right)^2_V}\over { \left({ \partial p \over  \partial V}\right)_T}}\; .
\end{equation}

and then,
\begin{eqnarray}\label{ceP}
c_P &=& c_V + { \left[f(\eta)-\eta f'(\eta)\right]^2 \over
f(\eta)+\frac13 \eta f'(\eta)} \cr \cr
&=& -\frac32 + {{4\, f(\eta)\left[ f(\eta)-\eta f'(\eta)\right] } \over
{f(\eta)+\frac13 \eta f'(\eta)}} \; .
\end{eqnarray}

The isothermal ($K_T$) and adiabatic ($K_S$) compressibilities take the form
\begin{eqnarray}\label{KT}
K_T &=& - { 1 \over V} \left({ \partial V \over  \partial p}\right)_T =
{V \over N\, T} {1 \over {f(\eta)+\frac13 \eta f'(\eta)}} \; ,\cr \cr
K_S &=& - { 1 \over V} \left({ \partial V \over  \partial p}\right)_S = 
{ c_V \over c_P} \; K_T \; .
\end{eqnarray}
It is then convenient to introduce the compressibilities
\begin{eqnarray}\label{kapa}
\kappa_T \equiv {NT \over V} \, K_T = {1 \over {f(\eta)+\frac13 \eta
f'(\eta)}} \quad \mbox{and} \quad \kappa_S \equiv{NT \over V} \, K_S = 
{ c_V \over c_P} \; \kappa_T \; ,
\end{eqnarray}
which are both of order one (intensive) in the $ N, \; L \to \infty $
limit with $ N/L $ fixed. 

\bigskip

The speed of sound $ v_s $ can be written as \cite{llmf}
\begin{equation}\label{defson}
v_s^2 = - {{ c_P \; V^2} \over {c_V \; N}} \left({ \partial p \over
\partial V}\right)_T = {V^2 \over N} \left[ {T \over N \, c_V}  \left({
\partial p \over  \partial T}\right)^2_V  - \left({ \partial p \over
\partial V}\right)_T \right] \; .
\end{equation}
where we used eq.(\ref{cpcv}) in the last step. 
Therefore,
\begin{equation}\label{vson}
{v_s^2 \over T} = { \left[f(\eta)-\eta f'(\eta)\right]^2\over 3
\left[f(\eta)-\eta f'(\eta)-\frac12 \right]} + f(\eta)+\frac13 \eta
f'(\eta)\; .
\end{equation}
The pressure $ p $ used in this calculation corresponds to the
pressure on the surface of the system. Hence, this is the speed of
sound on the surface of the system, this is different from the
speed of sound 
inside the volume since the ground state is inhomogeneous. We compute
the speed of sound as a function of the point in paper II.

\bigskip

We see that the large $ N $ limit of the self-gravitating gas is
special. Energy, free energy  and entropy are {\bf extensive} magnitudes
in the sense that they are proportional to the number of particles $ N
$ (for fixed $\eta$). 
They all depend on the variable $ \eta ={G \, m^2 N \over L \; T} $
which is to be kept fixed for the thermodynamic limit
($ N\to \infty $ and $ V \to \infty $) to exist.
Notice that  $ \eta $ contains the
ratio $ N/L = N \; V^{-1/3} $ which must be considered here an 
{\bf intensive variable}. Here, the presence of long-range gravitational
situations calls for this new intensive variable in the thermodynamic limit.

In addition, all physical magnitudes can be expressed in terms of a
single function of one variable: $ f(\eta) $.
\subsection{The diluted regime: $ \eta << 1$}

We can obtain the thermodynamic quantities as a series  in powers
of $ \eta $ just expanding the exponent in the integrand of $
\Phi_N(\eta) $ [eq.(\ref{FiN})]. 

To first order in $ \eta $ we get,
\begin{eqnarray}\label{etach}
\Phi_N(\eta) &=& \eta\; \int_0^1\ldots \int_0^1
\prod_{l=1}^N d^3r_l\;\; u({\vec r}_1,\ldots,{\vec r}_N)  
+ {\cal O}(\eta^2)  \cr \cr
&=& \frac12 \, \eta \; (N-1) \int_0^1 \int_0^1 { {d^3 r_1 \; d^3 r_2} \over
{ |{\vec r}_1 - {\vec r}_2|_a}} + {\cal O}(a^2)
+ {\cal O}(\eta^2)  \cr \cr
&=& 3(N-1)\; b_0 \; \eta + {\cal O}(\eta \, a^2)+ {\cal O}(\eta^2) \; .
\end{eqnarray}
where the coefficient $ b_0 $ is defined by eq.(\ref{b1b2}). 

To first order in $ \eta $ we see that the cutoff effect is
negligible $ \sim {\cal O}(a^2) $ [see (\ref{bcut})].

\bigskip

To second order in $ \eta $ we find from eq.(\ref{FiN}),
\begin{eqnarray}
e^{\Phi_N(\eta)}&=& \int_0^1\ldots \int_0^1
\prod_{l=1}^N d^3r_l\;\; e^{ \eta \; u({\vec r}_1,\ldots,{\vec r}_N)}
\cr \cr 
&=&1 + 3(N-1)\; b_0 \; \eta \cr \cr
&+& {\eta^2 \over 2 \, N^2} \left[ {N(N-1)(N-2)(N-3)\over 4} \int 
{d^3r_1 \;  d^3r_2 \; d^3r_3 \; d^3r_4 \over  |{\vec r}_1 - {\vec
 r}_2|  |{\vec r}_3 - {\vec r}_4| } \right.\cr \cr
&+& N(N-1)(N-2)\int {d^3r_1 \;  d^3r_2 \; d^3r_3 \over  |{\vec
r}_1 - {\vec r}_2|  |{\vec r}_1 - {\vec r}_3| }  \cr \cr 
&+& \left. {N(N-1) \over 2} \int { d^3r_1 \; d^3r_2 \over  |{\vec
r}_1 - {\vec r}_2|^2 }\right]+  {\cal O}(\eta^3, \eta \, a^2, \eta^2 \, a)\; .
\end{eqnarray}
where the coefficients in front of the integrals count the number of
combinations of particles yielding the same contribution. Using the
notation defined by eqs.(\ref{b1b2}) we get 
\begin{eqnarray}\label{etacu}
e^{\Phi_N(\eta)}&=& 1+3(N-1)\; b_0 \; \eta + \eta^2 \left[
{9(N-1)(N-2)(N-3)\over 2 N}  \; b_0^2 \right. \cr \cr 
&+& \left. {(N-1)(N-2)\over 2 N} \;
b_1 + {(N-1)\over 4 N} b_2  \right]+  {\cal O}(\eta^3, \eta \, a^2,
\eta^2 \, a)\; . 
\end{eqnarray}

Taking the log we get in the infinite $ N $ limit;
$$
\mbox{lim}_{N\to \infty}\frac{1}{N}\Phi_N(\eta) =  3 \, b_0 \eta
+\eta^2\left[ \frac12 b_1 - 18 \, b_0^2 \right] + {\cal O}(\eta^3)\; .  
$$
where we have now set $a=0$.

The cutoff effect is here again of order  $ \sim {\cal O}(a^2) $. It must be
noticed that the coefficient $ b_2 $ which has the stronger dependence
on the cutoff [see (\ref{bcut})] cancels out in the $ N = \infty $ limit.

We therefore find in the low density and the large $ N $ limit
using eqs.(\ref{estado}), (\ref{pVnT}) and (\ref{etach}):
\begin{equation}\label{petach}
{p V \over N T} = f(\eta) = 1 -  b_0 \; \eta -\eta^2\left[ \frac13 b_1
- 12 \,  b_0^2 \right] +  {\cal O}(\eta^3)\; .
\end{equation}

Furthermore, the speed of sound approaches for $ \eta \to 0 $  its
perfect gas value,
$$
{v_s^2 \over T} \; \;  {\buildrel{ \eta \downarrow 0}\over =}\; \;  \frac53
-\frac43\, b_0 \; \eta - \frac59 \,  \eta^2 \, \left[ b_1
- 36 \,  b_0^2 \right] + {\cal O}(\eta^3)\; .
$$
where we used eqs.(\ref{vson}) and (\ref{petach}).

As we see, there are no divergent contributions in $ \Phi_N(\eta) $ in
the zero cutoff limit to the second order in $ \eta $.  

At order three a logarithmically divergent integral appears in $
e^{\Phi_N(\eta)} $. Namely,
$$
{ \eta^3\over 3! \; N^3 } \frac12 N(N-1) \int { d^3r_1 \; d^3r_2 \over  |{\vec
r}_1 - {\vec r}_2|_a^3 } \sim { \eta^3 \over N} \log a
$$
This integral gives to $ f(\eta) $ and the other physical magnitudes a
contribution of the order
$$
{ \eta^3 \over N^2} \log a
$$
Therefore, such quantities can be {\bf safely neglected} for $ N \to \infty
$ and fixed (small) $a$ since $ f(\eta) $ is of order $ N^0 $ for $ N
\to \infty $. 

More generally, to the nth. order in $ \eta $ and $n>3$ the leading divergent
contribution to $ e^{\Phi_N(\eta)} $ for $ a \to 0 $ is of the form
$$
{ \eta^n\over n! \; N^n } \frac12 N(N-1) \int { d^3r_1 \; d^3r_2 \over  |{\vec
r}_1 - {\vec r}_2|_a^n } \sim { \eta^n \over  n! \; N^{n-2}} \; a^{3-n}
$$
This  gives to $ f(\eta) $ and the other physical magnitudes a
contribution of the order
$$
{ \eta^n \over n! \; N^{n-1}}\;  a^{3-n}
$$
As in the $ n = 3 $ case, such contributions are negligible in the $ N
\to \infty $ limit since we take it at fixed (small) $ a $.

\section{Microcanonical vs. Canonical Ensembles}

Let us  compare  the thermodynamical quantities computed in
the microcanonical and canonical ensembles in the $ N \to \infty $
limit keeping $ \xi $ and $ \eta $ fixed, respectively. 

We consider here the dilute limit where we dispose of  analytic
expressions. The Monte Carlo and mean field
results for the two ensembles will be compared in the next sections
and in paper II.

In the dilute limit, we have the expressions (\ref{microdilu}) and
(\ref{petach}) for the equation of state in the microcanonical and canonical
ensembles, respectively. We want to know whether they are or not equivalent.

Let us start from the microcanonical equation of state
(\ref{microdilu}). We have to express $ \eta $ in terms of $ \xi $ in
order to compare with the canonical equation of state (\ref{petach}). 

It follows from eqs.(\ref{tzi}), (\ref{EypV})  and (\ref{defeta}) that
$$
\eta = { g(\xi) \over \xi}
$$
Hence, for large $ \xi $ and small $ \eta $,
\begin{equation}\label{etamic}
 \eta  = \frac3{2\xi} -{ 9\, b_0\over 2 \xi^2} - {9 \over 4\,
\xi^3}\left(b_1  -  42 \, b_0^2 \right) + {\cal O}(\xi^{-4})
\end{equation}
and then,
\begin{equation}\label{tzimic}
{1 \over \xi} = \frac23 \, \eta \left[ 1 + 2 \, b_0 \, \eta - 2 \left(
10\, b_0^2 -\frac13 b_1\right)\, \eta^2 + {\cal O}(\eta^3) \right] \; .
\end{equation}
One easily sees that inserting eq.(\ref{tzimic}) in the microcanonical
equation of state (\ref{microdilu}) yields the canonical equation of
state (\ref{petach}) [up to  $ {\cal O}(\eta^3) = {\cal O}(\xi^{-3})
$]. 

Conversely, starting from the canonical ensemble, it follows from
eqs.(\ref{EypV}), (\ref{ecan}) and (\ref{petach}) that  
\begin{equation}\label{enecano}
{ E \over NT}  =  g(\xi)=3\left[  f(\eta) -\frac12\right]= \frac32 - 3\, b_0 \;
\eta - \eta^2 \, \left[ b_1 - 36 \,  b_0^2 \right] + {\cal O}(\eta^3)\; .
\end{equation}
and
$$
\xi = {3 \over \eta} \left[f(\eta) -\frac12\right] = \frac3{2\eta}\left[ 1 
- 2\, b_0 \; \eta - \frac23 \, \eta^2 \, \left( b_1 - 36 \,  b_0^2
\right) + {\cal O}(\eta^3) \right]\; . 
$$
We see that this relation is identical to eqs.(\ref{etamic}) and
(\ref{tzimic}) obtained in the microcanonical ensemble [up to  $ {\cal
O}(\eta^3) = {\cal O}(\xi^{-3}) $]. 

Inserting now eq.(\ref{etamic}) into the canonical equation of state
(\ref{petach}) yields the microcanonical equation of state
(\ref{microdilu}) [up to  $ {\cal O}(\eta^3) = {\cal O}(\xi^{-3}) $]. 

One verifies in the same way that all thermodynamical quantities
coincide to the same order in both ensembles.

In summary, the microcanonical and canonical ensembles yield the {\bf
same} results for $ N \to \infty $ to the orders $ \eta^0, \eta $ and
$ \eta^2 $ (or equivalently $ \xi^0, \xi^{-1} $ and $ \xi^{-3} $). 

\section{Monte Carlo calculations}

We have applied first the standard Metropolis algorithm\cite{montec} to the
self-gravitating gas in a cube of size $ L $ in the canonical
ensemble at temperature $ T $. We
computed in this way the pressure, the energy, the average density,
the potential energy fluctuations, the average particle distance
and the average squared particle distance as functions of $ \eta $. 
We implement the Metropolis algorithm changing at random the position of one
particle chosen at random. The energy of the configurations is
calculated performing the exact sums as in eq.(\ref{defu}). We used as
statistical weight for the Metropolis algorithm in the canonical ensemble,
$$
 e^{ \eta \; u({\vec r}_1,\ldots,{\vec r}_N)}\; ,
$$
which appears in the coordinate partition function (\ref{FiN}).

The number of particles $N$ went  up to $2000$. We introduced
a small short distance cutoff $ A = 10^{-4}L  - 10^{-8} L$ in the
attractive Newton's potential according to eq.(\ref{defva}). All
results in the gaseous phase were insensitive to the cutoff value. 
The partition function calculation turns to be much less sensible to
the short distance singularities of the gravitational force than
Newton's equations of motion for $N$ particles. That is, solving the
classical dynamics for $N$ particles interacting through gravitational
forces as well as solving the Boltzman equation including the $N$-body
gravitational interaction requires sophisticated algorithms to avoid
excessively long computer times \cite{deh}. As is clear, solving the $N$-body
classical evolution or the kinetic equations provides the
time-dependent dynamics and out of thermal equilibrium effects which
are out of the scope of our approach.  

\begin{figure}
\begin{turn}{-90}
\epsfig{file=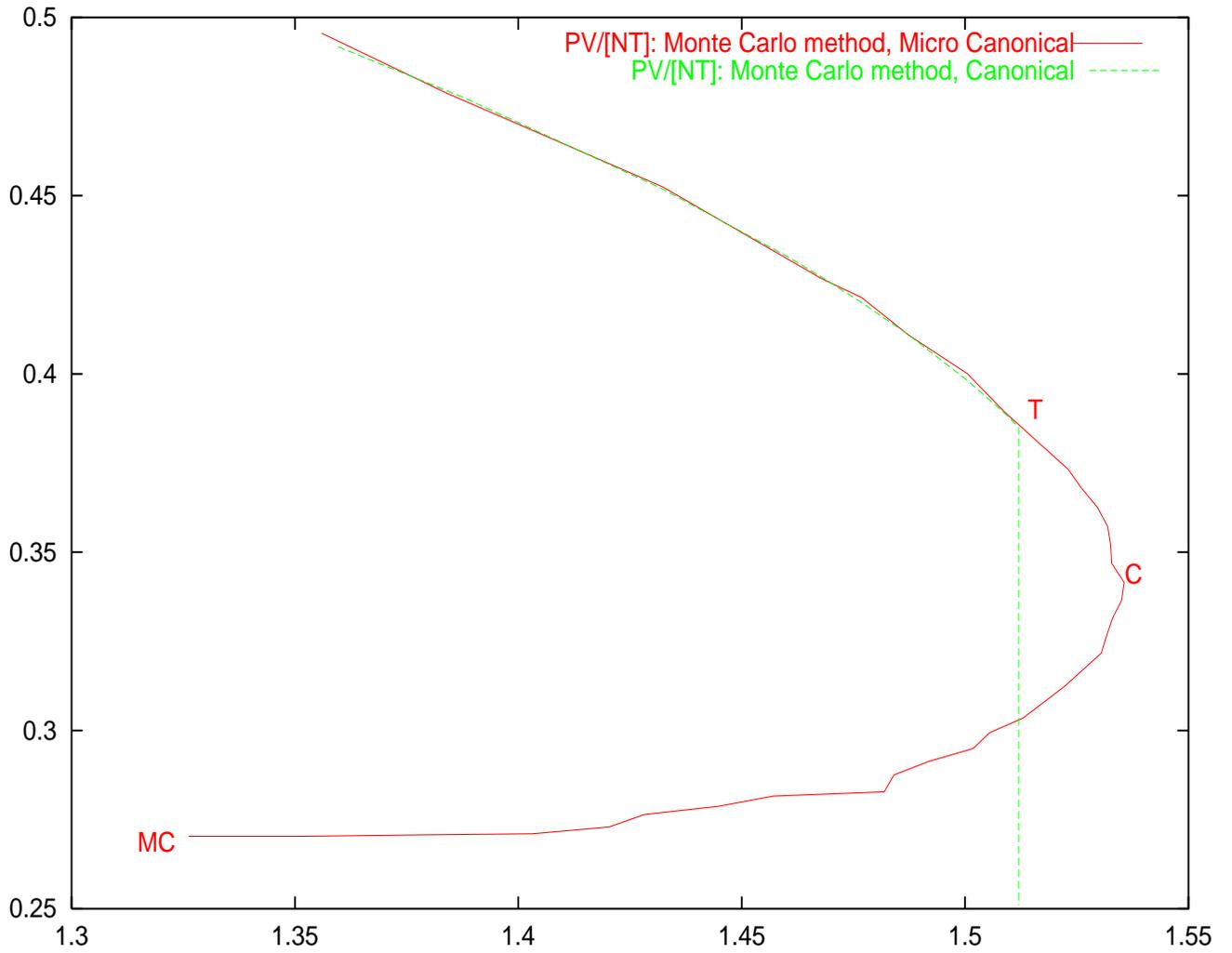,width=14cm,height=18cm} 
\end{turn}
\caption{ $f(\eta^R) = P V/[ N T]$ as a function of $ \eta^R $  by Monte
Carlo simulations for the microcanonical and canonical ensembles
($N=2000$). Both curves coincide within the statistical error till the point T.
\label{fig14}}
\end{figure}

\begin{figure}[t] 
\begin{turn}{-90}
\epsfig{file=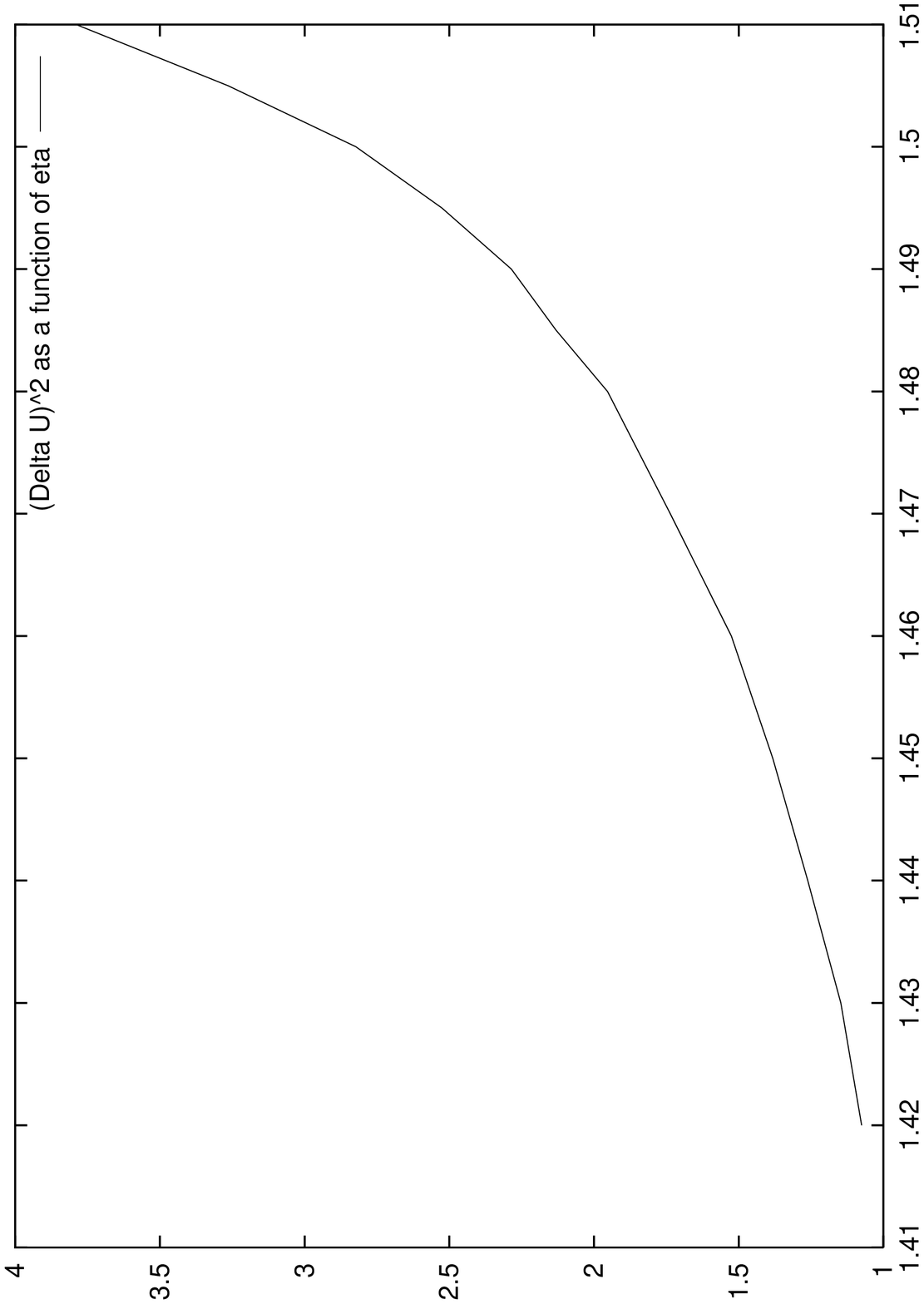,width=12cm,height=18cm} 
\end{turn}
\caption{ $ (\Delta U)^2 \equiv {{<U^2>-<U>^2}\over N \; T^2}= 3 \left[
f(\eta)-\eta \; f'(\eta) - 1 \right]  $ as a function of $ \eta $ in
the gaseous phase from Monte Carlo simulations with $2000$ particles
in the canonical ensemble. Recall that $ c_V = 3/2 +  (\Delta U)^2
$. \label{fig2}}  
\end{figure}

In the CE, two different phases show up: for $ \eta < \eta_T $ we have a
non-perfect gas and for $  \eta > \eta_T $ it is  a condensed
system with {\bf negative} pressure. The transition between  the two phases
is very sharp. This phase transition is associated with the Jeans
instability.

A negative pressure indicates that the free energy grows for
increasing volume at constant temperature [see
eq.(\ref{pres})]. Therefore, the system wants to contract sucking on
the walls.   

We plot in figs. 1 and 2, $ f(\eta) = pV/[NT] $ and $ (\Delta U)^2 $ as
functions of $ \eta $, respectively. 

We find that  for small $ \eta $,  the Monte Carlo results for  $ pV/[NT] $
well reproduce the analytical formula (\ref{petach}).  $ pV/[NT] $
monotonically decreases with  $ \eta $. 

In the Monte Carlo simulations the phase transition to the condensed
phase happens for $ \eta = \eta _T $ slightly below $ \eta_C $.
For $ N = 2000 $ we find $ \eta_T \sim 1.515 $.  
For $ \eta _T < \eta < \eta_C $, the gaseous phase can only exist as a
metastable state.

The average distance between particles $ <r> $  and the average
squared distance between particles $ <r^2> $ monotonically decrease
with  $ \eta $. When the gas collapses at   $ \eta_T \; , <r> $
and $ <r^2> $ exhibit a sharp decrease.  

The values of $ pV/[NT] , \;  <r> $ and $ <r^2> $ in the condensed phase are
independent of the cutoff for $ a < 10^{-5} $. The Monte Carlo results
in this condensed phase can be approximated for 
$ \eta > 2 $ 
as 
\begin{equation}\label{pcolap}
{pV \over NT} = f(\eta) \simeq  1 -  K \; \eta \quad ,
\quad  <r> \simeq 0.016  \; .
\end{equation}
where $ K \simeq 14 $. 

\bigskip

Since $ f(\eta) $ has a jump at the transition, the Gibbs free energy
$ \Phi $ is discontinuous and we have a phase
transition of the {\bf zeroth} order. We find from eq.(\ref{gibbs})
\begin{equation}\label{deltaG}
{\Phi(\mbox{collapse})-\Phi(\eta_T)  \over N \, T} =
f(\mbox{collapse}) - f(\eta_T) \simeq -21  < 0 \; . 
\end{equation}

We can easily compute the latent heat of the transition per particle ($q$)
using the fact that the volume $V$ stays constant. Hence, $ q = \Delta
E/N $ and we obtain from eq.(\ref{ecan})
\begin{equation}\label{qsobreT}
{q \over T} = {E(\mbox{collapse}) -E(\eta_T)  \over N \, T} =3 \; \left[
f(\mbox{collapse}) - f(\eta_T) \right] 
\simeq 2 - 3  \, K \; \eta_T\simeq -62  < 0 \; . 
\end{equation}
This phase transition is different from the usual phase transitions
since the two phases cannot coexist in equilibrium as their pressures
are different.

Eq.(\ref{pcolap}) can be understood from the general treatment in
sec. III as follows. We have from eqs.(\ref{estado})-(\ref{fiprima}) 
\begin{equation}\label{Feta}
f(\eta) = 1 - {\eta \over 3} < {1 \over r}> \; .
\end{equation}
The Monte Carlo results indicate that $  < {1 \over r}> \simeq 42 $ is
approximately constant in the collapsed region as well as $ <r> $ and
$ <r^2> $.  Eq.(\ref{pcolap}) thus follows from eq.(\ref{Feta}) using
such value of $ < {1 \over r}> $. 

\bigskip

The behaviour of $ pV/[NT] $ near  $ \eta_C $ in the gaseous phase 
can be well reproduced by 
\begin{equation}\label{pecrit}
{pV \over NT}= f(\eta) \;\; {\buildrel{ \eta \uparrow \eta_C}\over =}\;
 f_C + A \; \sqrt{\eta_C -\eta} 
\end{equation}
where $  f_C \simeq 0.316, \; A \simeq 0.414 $ and  $ \eta_C \simeq 1.540 $.

In addition, the
behaviour of $ (\Delta U)^2 $ in the same region is well reproduced by
\begin{equation}\label{flUc}
(\Delta U)^2\; \; {\buildrel{ \eta \uparrow \eta_C}\over =}\;\; C  + 
{D \over \sqrt{\eta_C-\eta} }
\end{equation}
with $ C \simeq -1.64 $  and $ D \simeq 0.901 $. 
[Notice that for finite $ N , \; (\Delta U)^2 $ will be finite albeit
very large at the phase transition]. Eq.(\ref{delu}) relating $
f(\eta) $ and $ (\Delta U)^2 $ is satisfied with reasonable
approximation. 

We thus find a critical region just below $ \eta_C $ where the energy
fluctuations tend to infinity as $ \eta \uparrow \eta_C $.

The point $ \eta_T $ where the phase transition actually takes place
in the Monte Carlo simulations is at $ \eta_T \simeq 1.51 < \eta_C
$. This value for $ \eta_T $ is close to the point $  \eta_0 $ where the
isothermal compressibility $ \kappa_T $ diverges (see sec. VII). They
are probably the same point. 

Since Monte Carlo simulations are like real experiments, we conclude
that the gaseous phase extends from $ \eta = 0 $ till $ \eta = \eta_T
$  in the CE and {\bf not} till $ \eta = \eta_C $. 
Notice that in the literature based on the hydrostatic
description of the self-gravitating gas \cite{sas,pad,HK,bt}, only the
instability at $ \eta = \eta_C $ is discussed whereas the
singularities at $ \eta = \eta_0 $ are not considered.

\bigskip

We then performed Monte Carlo calculations in the microcanonical
ensemble where the coordinate partition function is given by eq.(\ref{defw}). 
We thus used  
$$
\left[\xi + {1 \over N}\, u({\vec r}_1, \ldots, {\vec r}_N) \right]^{3N/2
-1}\theta\left[\xi + 
{1 \over N} \, u({\vec r}_1, \ldots, {\vec r}_N) \right]\; ,
$$
as the statistical weight for the Metropolis algorithm. 

The MCE and CE Monte Carlo results coincide up to the
statistical error for $0 < \eta <\eta_T $, that is for $ \infty >
\xi > \xi_T \simeq -0.19 $. In the MCE the gas does not clump at $\eta
= \eta_C $ (point $C$ in fig. \ref{fig14}) and the specific heat
becomes negative 
between the points $C$ and $MC$. In the MCE the gas does clump at
$ \xi \simeq -0.52 \; , \; \eta^T_{MC} \simeq 1.33 $ (point $MC$ in
fig. \ref{fig14}) increasing  {\bf both its temperature and pressure
discontinuously}. We find from the Monte Carlo data that the
temperature increases by a factor $ 2.4 $ whereas the pressure
increases by a factor $ 3.6 $ when the gas clumps. The transition
point $ \eta^T_{MC} $ in 
the Monte Carlo simulations is slightly to the right of the critical
point $ \eta_{MC} $  predicted by mean field theory. The mean field
yields for the sphere $ \eta_{MC} = 1.2598\ldots $. 

In ref.\cite{katz} finite $N$ corrections to the critical point $
\eta_{MC} $ are computed in mean field for the sphere. This finite $N$
corrections shift $ \eta_{MC} $ by $+3.3\%$ for $N=2000$. Since, $
\eta^T_{MC} $ differs from $ \eta_{MC} $ by $ +5.6\%$, $ \eta^T_{MC} $
and $ \eta_{MC} $ are probably  {\bf different} critical points.

\bigskip

As is clear, the domain between $C$ and $MC$ cannot
be reached in the CE since $ c_V > 0 $ in the CE as shown by eq.(\ref{cV}).  

We find an excellent agreement between the Monte Carlo  and Mean
Field (MF) results
(both in the MCE and CE). (This happens although the  geometry for the MC
calculation is cubic while it is spherical for the MF). The points
where the collapse phase transition  occurs ($\eta_T$ and $\eta^T_{MC}$) 
slowly increase with the number of particles $N$.

We verified that the Monte Carlo results in the gaseous phase  
($ \eta < \eta_T $) are cutoff independent for $ 10^{-3} \geq a \geq 10^{-7}
$. 

\bigskip

As for the CE, the Gibbs free energy is discontinuous at the
transition in the MCE. The transition is then of the zeroth order.
We find from eq.(\ref{gibbs})
$$
{\Phi(\mbox{collapse})-\Phi(\eta_T)  \over N \, T_{gas}} =
 {T_{coll} \over T_{gas}}\, f(\mbox{collapse}) - f(\eta_T) \simeq
 0.7  > 0 \; .  
$$
where we used the numerical values from the Monte Carlo simulations.
Notice that the Gibbs free energy {\bf increases} at the MC transition
whereas it decreases at the C transition [see eq.(\ref{deltaG})]. 

Here again
the two phases cannot coexist in equilibrium since their pressures and
temperatures are different.

\bigskip

\begin{figure}[t] 
\begin{turn}{-90}
\epsfig{file=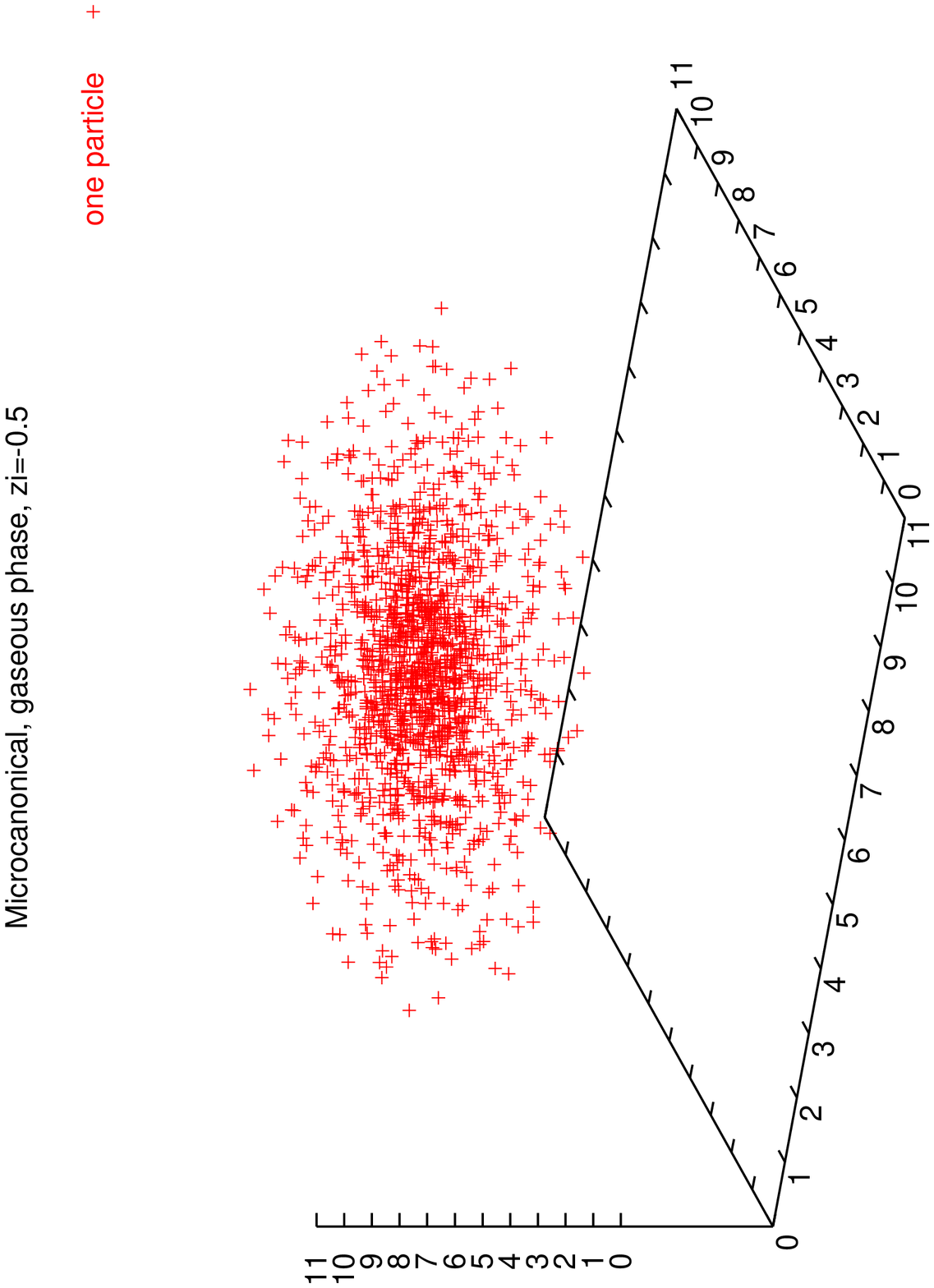,width=12cm,height=18cm} 
\end{turn}
\caption{ Average particle distribution in
the gaseous phase from Monte Carlo simulations with $2000$ particles
in the microcanonical ensemble for $\xi = - 0.5, \; \eta = 1.38, \;
pV/[NT] = 0.277$.  \label{gasmc}}  
\end{figure}

\begin{figure}[t] 
\begin{turn}{-90}
\epsfig{file=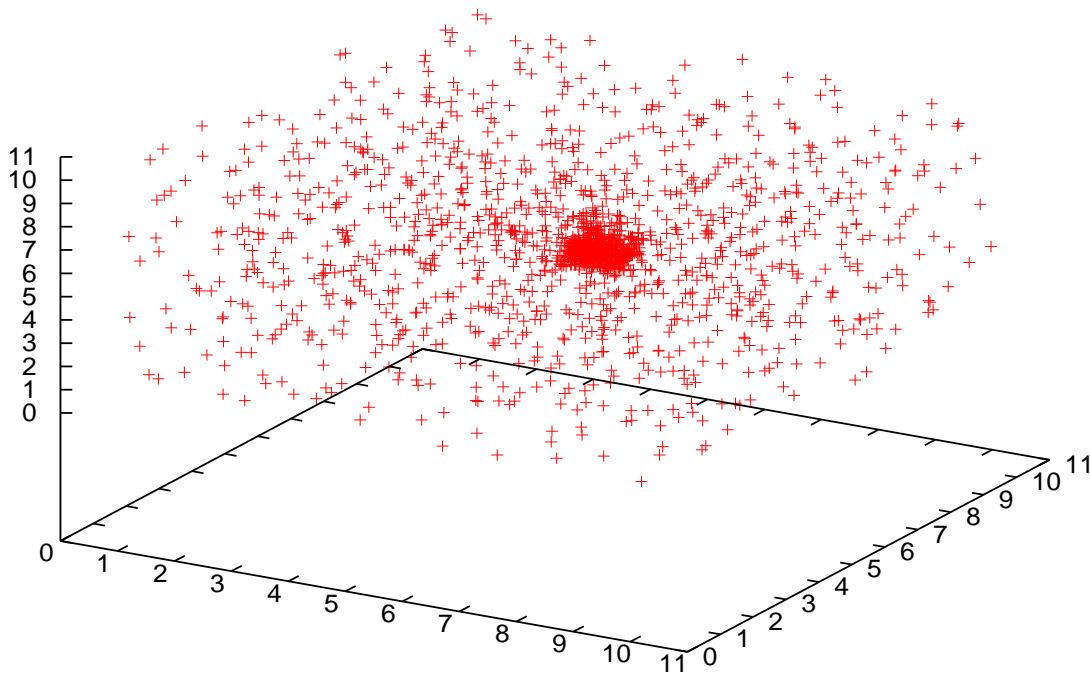,width=12cm,height=18cm} 
\end{turn}
\caption{  Average particle distribution in
the collapsed phase from Monte Carlo simulations with $2000$ particles
in the microcanonical ensemble for $\xi = - 0.6, \; \eta = 0.43, \;
pV/[NT] = 0.414 $.  \label{colmc}}  
\end{figure}

We display in figs. \ref{gasmc}-\ref{colmc} the average particle
distribution from 
Monte Carlo simulations with $2000$ particles in the microcanonical
ensemble at both sides of the gravothermal catastrophe, i. e. $ \eta =
\eta_{MC} $. Fig. \ref{gasmc} corresponds to the gaseous phase and
fig. \ref{colmc} to the collapsed phase. The inhomogeneous particle
distribution is clear in fig. \ref{gasmc} whereas fig. \ref{colmc}
shows a dense collapsed core surrounded by a halo of particles. 

The different nature of the collapse in the CE and in the MCE can be
explained using the virial theorem [see eq.(\ref{virial2})]
$$
{p \, V \over N \, T} = 1 + { U \over N \, T} \; .
$$
When the gas collapses in the CE the particles get very close and $ U
$ becomes large and negative while $ T $ is fixed. Therefore, $ p \, V
\over N \, T $ may become large and negative as it does.

We can write the virial theorem also as,
$$
p \, V - \frac12 \, NT = \frac13 \, E\; .
$$
When the gas is near the point MC, $ E < 0 $ is fixed and we have $ T
> 0 $. Therefore, $ p \, V \over N \, T $ as well as $ U = E - 3\, N \,
 T / 2 $ cannot become large and negative as in the CE collapse.
This prevents the distance between the particles to
decrease. Actually, the Monte Carlo simulations show that $ <r> $ {\bf
increases} by $ 18\% $ when the gas collapses in the MCE. 

\begin{figure}[t] 
\begin{turn}{-90}
\epsfig{file=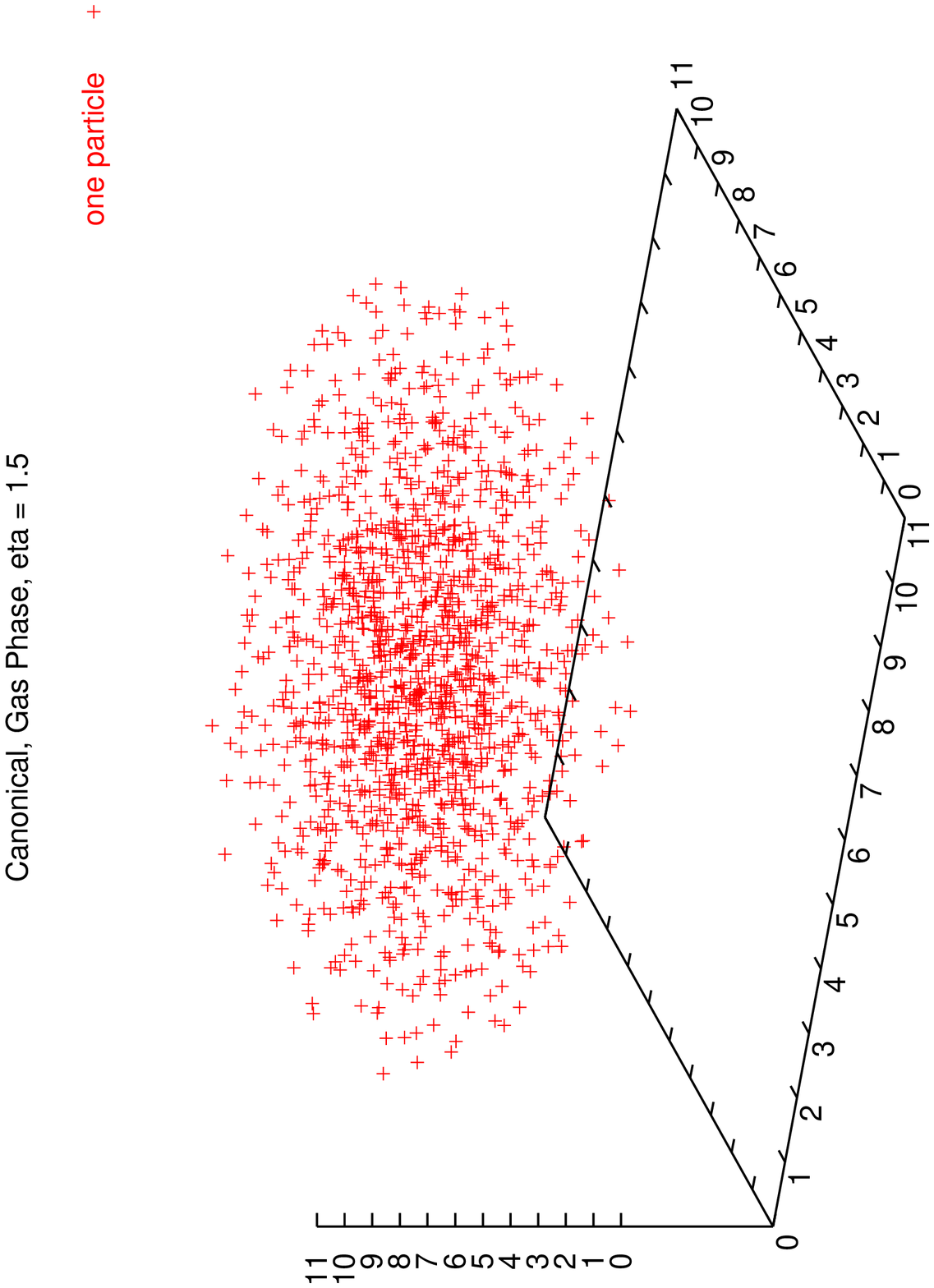,width=12cm,height=18cm} 
\end{turn}
\caption{Average particle distribution in the gaseous phase from Monte
Carlo simulations in the 
canonical ensemble for $ \eta = 1.5 $ and $ N = 2000 $ 
\label{gasc}} 
\end{figure}

\begin{figure}[t] 
\begin{turn}{-90}
\epsfig{file=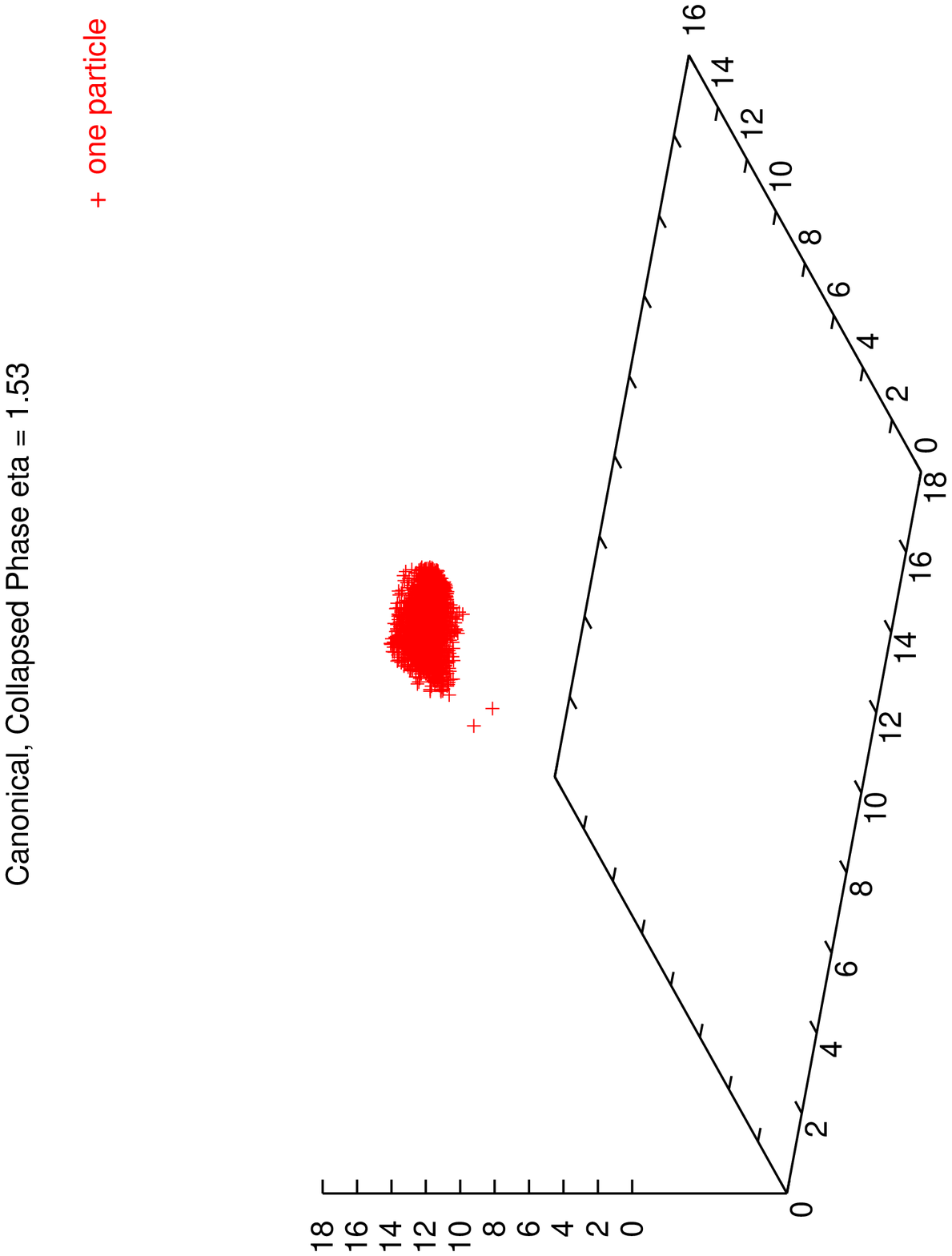,width=12cm,height=18cm} 
\end{turn}
\caption{  Average particle distribution in
the collapsed phase from Monte Carlo simulations with $2000$ particles
in the canonical ensemble for $ \eta = 1.53, \;
pV/[NT] = -14.44 $.  \label{colc}}  
\end{figure}

Figs. \ref{gasc} and \ref{colc} depict the average particle
distribution from  Monte Carlo simulations with $2000$ particles in
the canonical ensemble at both sides of the collapse critical point,
i. e. $ \eta = 
\eta_{C} $. Fig. \ref{gasc} corresponds to the gaseous phase and
fig. \ref{colc} to the collapsed phase. The inhomogeneous particle
distribution is clear in fig. \ref{gasc} whereas fig. \ref{colc}
shows a dense collapsed core surrounded by a very little halo of particles.

Notice that the collapsed phases are of different nature in the CE and
MCE. The core is much tighter and the halo much smaller  in the CE
than in the MCE.

Figs. 3 and 5 depict the average particle distribution for the gaseous
phase in the MCE and the CE, respectively. In this phase, the MC
simulations  give identical descriptions for large $N$ in both
ensembles. [This important point will be further demostrated in
sec. VI by functional integral methods]. The average configurations in
figs. 3 and 5 describe a self-gravitating gas in thermal equilibrium
within a {\bf cube}. We may call it the {\bf isothermal cube} by
analogy with the well known isothermal sphere\cite{emden}-\cite{bt}.

\section{Mean Field Approach}

Both in the microcanonical and the canonical ensembles the coordinate
partition functions are given by $3N$-uple integrals [eqs.(\ref{defw})
and (\ref{FiN}), respectively]. In the $ N \to \infty $ limit both $3N$-uple
integrals can be recasted as functional integrals over the continuous
particle density as we see below.

\subsection{The Canonical Ensemble}

We now recast the coordinate partition function $ e^{\Phi_N(\eta)} $ 
in the canonical ensemble
given by eq.(\ref{FiN}) as a functional integral in the thermodynamic limit. 
\begin{eqnarray}\label{zcanmf}
e^{\Phi_N(\eta)} &\buildrel{ N>>1}\over=& \int\int D\rho\; d{\hat a} \;
e^{-N s_C[\rho(.),\hat a,\eta]} \\ \cr
s_C[\rho(.),\hat a,\eta] &=& -{\eta \over2} \int  {d^3x  \; d^3y \over |{\vec
x}-{\vec y}|} \;\rho({\vec x}) \; \rho({\vec y}) + \int d^3x \; \rho({\vec
x}) \; \log\rho({\vec x}) - i{\hat a}\left(\int  d^3x \,
\rho({\vec x}) - 1 \right) \;.\nonumber 
\end{eqnarray}
where we used the coordinates $ \vec x $ in the unit volume.
The first term  is the potential energy, the second term is the functional
integration measure for this case (see appendix A). Here $ N \;
\rho({\vec x}) $ stands for the density of particles. 

The integration over $ \hat a $ enforces the number of particles to be
exactly $ N $:
\begin{equation}\label{vincu}
\int  d^3x \; \rho({\vec x}) = 1
\end{equation}
That is, in the coordinates $ \vec q $ (running from $ 0 $ to $ L $),
the density of particles is
$$
{N \over L^3} \; \rho({\vec q}) \quad {\rm with} \quad\int  d^3q \;{N
\over L^3} \; \rho({\vec q}) = N \; .
$$

\subsection{The Microcanonical Ensemble}

Let us express the coordinate partition function in the microcanonical
ensemble $ w(\xi,N) $ defined by eq.(\ref{defw}) in terms of the
coordinate partition function in the canonical ensemble $
e^{\Phi_N(\eta)} $  defined by eq.(\ref{FiN}). In order to do that we
use the Fourier expansion \cite{gel}
\begin{equation}\label{traFou}
x^{\lambda} \; \theta(x) = { \Gamma(\lambda+1) \over 2 \pi}
\int_{-\infty}^{+\infty} e^{i \, \omega \, x} { d\omega \over (i
\omega)^{\lambda+1}} 
\end{equation}
We thus find from eqs.(\ref{defw}), (\ref{FiN}) and (\ref{traFou})
that
\begin{eqnarray}\label{cancan}
w(\xi,N) &=& \Gamma\left({3 N \over 2}\right) \int_{-\infty}^{+\infty} {
d\omega \over 2 \pi} \; e^{i \omega \xi + \Phi_N(i \omega /N)
-{3N\over 2}\log(i\omega)}\cr \cr
 &=& N \, \Gamma\left({3 N \over 2}\right) \int_{\gamma} { d \eta
\over 2 \pi i} \; e^{N \, \eta \, \xi + \Phi_N(\eta) -{3N\over
2}\log(N\, \eta)}
\end{eqnarray}
where we introduced the integration variable $ \eta \equiv i \omega /N
$ and where $ \gamma $ is an upward integration contour parallel to the
imaginary $ \eta $ axis. 
Using Stirling's approximation for the $ \Gamma $ function we find for
$ N \gg 1 $ up to irrelevant constants
$$ 
w(\xi,N) \buildrel{ N>>1}\over=  \int_{\gamma} { d \eta \over 2 \pi
i} \; e^{N \, \eta \, \xi + \Phi_N(\eta) -{3N\over 2}\log\eta }
$$ 
Now, inserting the functional integral representation (\ref{zcanmf}) for the
coordinate canonical partition function yields,
\begin{equation}\label{ifmcan}
w(\xi,N) \buildrel{ N>>1}\over= \int\int D\rho\; d{\hat a} \;{ d
\eta \over 2 \pi i}\; 
e^{N\left[  \eta \, \xi  -{3\over 2}\log\eta - s_C[\rho(.),\hat a,\eta]\right]}
\end{equation}
We  thus find a functional integral representation in the microcanonical
ensemble analogous to the canonical representation eq.(\ref{zcanmf})
but with an extra integration (over $ \eta $) that constrains the
value of the energy. 

The `effective action' in the microcanonical ensemble takes thus the
form,
\begin{equation}\label{accefmc}
s_{MC}[\rho(.),\hat a, \eta] = {3\over 2}\log\eta-\eta \, \xi
-{\eta \over2} \int  {d^3x  \; d^3y \over |{\vec
x}-{\vec y}|} \;\rho({\vec x}) \; \rho({\vec y}) + \int d^3x \; \rho({\vec
x}) \; \log\rho({\vec x}) - i{\hat a}\left(\int  d^3x \,
\rho({\vec x}) - 1 \right)
\end{equation}

\subsection{The Grand Canonical Ensemble}

The partition function in the grand canonical ensemble can be written
as
\begin{equation}\label{defpgc}
{\cal Z}_{GC}(z,T)= \sum_{N=0}^{\infty} z^N \;{\cal Z}(N,T) \; ,
\end{equation}
where $ z = e^{\mu \over T} $ stands for the fugacity and $ {\cal
Z}(N,T) $ is the partition function in the canonical ensemble given by
eqs.(\ref{fp}) and (\ref{fp2}).

As shown in ref.\cite{prd}, this grand canonical partition function
can be expressed as a functional integral
\begin{equation}\label{zfi}
{\cal Z}_{GC}(z,T) =  \int\int\;  {\cal D}\Phi\;  e^{ {1\over{T_{eff}}}\;
\int_V d^3q \left[ \frac12\Phi \; \nabla^2\Phi \; + M^2 \; e^{\Phi({\vec
q})}\right]}\; ,  
\end{equation}
where
\begin{equation}\label{muyT}
M^2 = \sqrt{2\; T \over {\pi}}\; z\; G \, m^{7/2} 
\quad , \quad T_{eff} = 4\pi \; {{G\; m^2}\over {T}} \; .
\end{equation}
Notice that the representation (\ref{zfi}) is exact while the
functional integral representations in the microcanonical and
canonical ensembles only apply for large number of particles. 

Rewriting eq.(\ref{zfi}) in terms of the dimensionless variables
(\ref{variar}) yields for the exponent 
$$
{1\over{T_{eff}}}\;
\int_V d^3q \left[ \frac12\Phi \; \nabla^2_q\Phi \; + \mu^2 \; e^{\Phi({\vec
q})}\right] = {L\over T_{eff} }\;
\int_0^1 d^3x \left[ \frac12\Phi \; \nabla^2_r\Phi \; + \zeta^2 \;
e^{\Phi({\vec x})}\right]
$$
where $ \zeta \equiv M \; L $ is of the order one ($ L^0 $), since $
M^2 \sim z = e^{\mu \over T} \sim L^{-2} $ [see eq.(\ref{poqui})].

Since the exponent in the functional integral (\ref{zfi}) is
proportional to $ L $, the large volume limit is dominated by the
stationary points (mean field approximation)
\begin{equation}\label{ecdelgc}
\nabla^2_r\Phi_s({\vec x}) \; + \zeta^2 \; e^{\Phi_s({\vec x})} = 0 \; .
\end{equation}
We expand around the saddle point $ \Phi_s({\vec x}) $ changing to a
new functional integration $ Y({\vec x}) $ variable as follows,
\begin{equation}\label{flucgc}
\Phi({\vec x}) =  \Phi_s({\vec x}) + Y({\vec x}) \; .
\end{equation}
Keeping in eq.(\ref{zfi}) quadratic terms in $ Y(.) $ yields,
\begin{equation}\label{gausgc}
{\cal Z}_{GC}(z,T) = e^{{L\over T_{eff} }\;
\int_0^1 d^3x \left[ \frac12\Phi_s \; \nabla^2_r\Phi_s \; + \zeta^2 \;
e^{\Phi_s({\vec x})}\right]} \;  \int\int\;  {\cal D}Y\;e^{ {L\over 2
\; T_{eff} }\;\int_0^1 d^3x \; \left[ Y \nabla^2 Y + \zeta^2 \, Y^2
\;e^{\Phi_s({\vec x})} \right]}\left[ 1 + {\cal O}\left( {1 \over
L}\right) \right]   
\end{equation}
where the Gaussian integral over $ Y(.) $ gives a factor of order $
L^0 $ [see paper II].

\bigskip

We recall that the saddle point method applies while all eigenvalues
of the quadratic form in the exponent of eq.(\ref{gausgc}) are
positive. Therefore, the determinant of the quadratic fluctuations is
positive. The determinant vanishing or changing sign indicates the
presence of zero or negative eigenvalues. In such a case the system is
no more on a stable phase but on a metastable or unstable phase. The
free energy gets an imaginary part in such metastable or unstable
situations. 

\bigskip

The average number of particles in the grand canonical ensemble is given by
$$
{\bar N} = {1 \over {\cal Z}_{GC}} \sum_{N=0}^{\infty}N z^N \;{\cal Z}(N,T) 
= \left. { \partial \log {\cal Z}_{GC} \over\partial \log z}\right|_{V,T} \; .
$$
We thus find in the mean field approximation,
$$
{\bar N} = { L \; \zeta^2 \over T_{eff} } \int_0^1 d^3x \; e^{\Phi_s({\vec x})}
$$

Therefore, using this and eq.(\ref{muyT}) we can express $  \zeta^2 $
in terms of $ \eta $ where we denote $ \bar N $ as $ N $ to avoid
cluttering of notation,
\begin{equation}\label{zeda}
 \zeta^2 = { 4\pi \eta \over \int_0^1 d^3x \;  e^{\Phi_s({\vec x})}}
 \; ,
\end{equation}
and the fugacity results
\begin{equation}\label{zgc}
z = {N \over L^3 } \left({2\pi\over mT  }\right)^{3/2}{ 1 \over
\int_0^1 d^3x \;  e^{\Phi_s({\vec x})}} \; .
\end{equation}
We again see that $ z \sim L^{-2} $ in the GCE. 

Integrating eq.(\ref{ecdelgc}) over the unit volume yields
\begin{equation}\label{intsgc}
\int {\vec \nabla} \Phi_s({\vec x}) \cdot d{\vec s} = - 4\pi \eta
\end{equation}
where we used eq.(\ref{zeda}).

\bigskip

We find for the free energy\cite{llms},
\begin{equation}\label{efegc}
F= -T \log {\cal Z}_{GC} + NT \; \log z =F_0 + {NT \over 2} K(\eta) -
NT \, \log C(\eta) + {\cal O}(N^0) \; ,
\end{equation}
where we used the grand canonical partition function (\ref{gausgc})
evaluated at the stationary point,
\begin{equation}\label{Zgcmf}
\log {\cal Z}_{GC} = N \left[ 1 - \frac12 \, K(\eta)\right] \; ,
\end{equation}
and $ z $ is given by eq.(\ref{zgc}) with
\begin{equation}\label{defKC}
K(\eta) \equiv { \int_0^1 d^3x \; \Phi_s({\vec x}) \;  e^{\Phi_s({\vec
x})} \over C(\eta)} \quad \mbox{and} \quad 
C(\eta) \equiv \int_0^1 d^3x \;  e^{\Phi_s({\vec x})} \; .
\end{equation}
$ F_0 $ is given by eq.(\ref{Fcero}).

We easily calculate the mean value of the potential energy in the mean
field approximation
\begin{equation}\label{ugc}
<U> = - T \; { \partial \log {\cal Z}_{GC} \over \partial \log G}= 
-{NT \over 2} K(\eta)
\end{equation}

Combining the two expressions for the entropy
\begin{equation}\label{relter}
S = {E-F \over T} \quad \mbox{and} \quad  S = - \left({ \partial F \over
\partial T}\right)_V \; ,
\end{equation}
yields
\begin{equation}\label{entgc}
S = S_0 - N\left[  K(\eta) - \log C(\eta)\right]
\end{equation}
and the first order differential equation
\begin{equation}\label{ecKC}
\eta \, K'(\eta) +  K(\eta)  = 2 \, \eta {d \over d \eta} \log
C(\eta) \; .
\end{equation}
The boundary conditions $ K(0) = 0, \; C(0) = 1 $ ensure the ideal gas
limit $ \eta = 0 $.

The pressure takes the form,
\begin{equation}\label{prgc}
P = - \left({ \partial F \over  \partial V}\right)_T = {N T \over V}
\left[ 1 + { \eta \over 3} \; \left( \frac12 K'(\eta) - {d \over d \eta} \log
C(\eta) \right) \right] + {\cal O}(N^0)\; .
\end{equation}
These equations guarantee in addition that the virial theorem
(\ref{virial2}) holds. 

\subsection{Saddle point evaluation in the canonical ensemble}

The functional integral in  eq.(\ref{zcanmf}) is dominated for large $N$ by
the extrema of the `effective action' $ s_C[\rho(.),\hat a,\eta] $,
that is, the solutions of the stationary point equation
\begin{equation}\label{eqrho}
\log \rho_s({\vec x}) - \eta \int{ d^3y \; \rho_s({\vec y})\over |{\vec
x}-{\vec y}|} =a_s \; ,
\end{equation}
$ a = i{\hat a} $ is a Lagrange multiplier enforcing 
the constraint (\ref{vincu}).

Applying the Laplacian and setting $\phi({\vec x}) \equiv
\log\rho_s({\vec x})$ yields, 
\begin{equation}\label{puntoest}
\nabla^2\phi({\vec x}) +   4\pi \eta \; e^{\phi({\vec x})} = 0 \; ,
\end{equation}
This equation is scale covariant \cite{prd}. That is, if $ \phi({\vec
x}) $ is a solution of eq.(\ref{puntoest}), then
\begin{equation}\label{coves}
\phi_{\lambda}({\vec x}) \equiv \phi(\lambda{\vec x}) +\log\lambda^2
\end{equation}
where $ \lambda $ is an arbitrary constant
is also a solution of eq.(\ref{puntoest}). For spherically symmetric
solutions this property  can be found in ref.\cite{chandra}. 

Integrating eq.(\ref{puntoest}) over the unit volume and using the
constraint (\ref{vincu}) yields
\begin{equation}\label{vincu2}
\int {\vec \nabla} \phi({\vec x}) \cdot d{\vec s} = - 4\pi \eta
\end{equation}
where the surface integral is over the boundary of the unit volume.

Comparing eqs.(\ref{ecdelgc})-(\ref{intsgc}) with (\ref{puntoest}) and
(\ref{vincu2}) shows that the grand canonical and canonical stationary
points are related by 
\begin{equation}\label{cangcan}
\Phi_s({\vec x}) = \phi({\vec x}) + \log C(\eta) \; .
\end{equation}
Eq.(\ref{gausgc}) can then be written as 
\begin{eqnarray}\label{gausgc2}
&&{\cal Z}_{GC}(z,T) = e^{{N\over  4\pi\eta }\; \left\{
\int_0^1 d^3x \left[ \frac12\phi \; \nabla^2_r\phi \; + 4\pi\eta \;
e^{\phi({\vec x})}\right] - 2 \pi\eta \log C(\eta) \right\} } \times \cr\cr
&&\int\int\;
{\cal D}Y\;e^{ {N \over 8\pi\eta}
\;\int_0^1 d^3x \; \left[ Y \nabla^2 Y +  4\pi\eta \, Y^2
\;e^{\phi({\vec x})} \right]}\left[ 1 + {\cal O}\left( {1 \over
N}\right) \right]   
\end{eqnarray}
where we used eqs.(\ref{defeta}), (\ref{muyT}), (\ref{zeda}) ,
(\ref{vincu2}) and (\ref{cangcan}). 

\bigskip

We have taken the zero cutoff limit in
eqs.(\ref{eqrho})-(\ref{puntoest}). The mean field equations turn to
be {\bf finite} with {\bf regular} solutions in such limit. This can
be understood from our perturbative calculation in sec. III.A. All
potentially divergent contributions at zero cutoff are suppressed by
factors $ 1/N^2 $ and therefore disappear in the $ N = \infty $
limit. Hence one can set the  cutoff to zero in the mean field
approximation. 

\bigskip

In order to evaluate the functional integral in eq.(\ref{zcanmf}) by the
saddle point method we change the functional integration variable as follows, 
\begin{equation}\label{camvar}
\rho({\vec x}) = \rho_s({\vec x}) + Y({\vec x}) \quad , \quad a = a_s
+ y_0
\end{equation}
where $ \rho_s({\vec x}) $ and $ a_s $ obey eq.(\ref{eqrho}). We can
expand the exponent to second order as
\begin{equation}\label{s2}
s_C[\rho(.),a,\eta] - s_C[\rho_s(.),a_s,\eta] = s^{(2)}_C[Y(.),y_0] +
{\cal O}\left( Y^3, Y^2 \, y_0\right) 
\end{equation}
where we use that
$$
\left. { \delta s_C \over \delta \rho({\vec x})}\right|_{\rho=\rho_s,\,
a=a_s}= 0 \quad , \quad \left. {\partial s_C \over \partial
a}\right|_{\rho=\rho_s,\, a=a_s}= 0  
$$
and
$$
\left. \left. s^{(2)}_C[Y(.),y_0] \equiv \frac12  \int  d^3x  \; d^3y \;
Y({\vec x})\; 
Y({\vec y}) \; { \delta^2 s_C \over \delta \rho({\vec x})\delta
\rho({\vec y}) }\right|_{\rho=\rho_s,\, a=a_s} + y_0 \; \int  d^3x \;
Y({\vec x}) \; { \delta^2 s_C \over \delta \rho({\vec x}) \partial
a}\right|_{\rho=\rho_s,\, a=a_s}  \; .
$$
Notice that 
$$
{\partial^2 s_C \over \partial a^2}= 0 \; .
$$
We evaluate explicitly the second derivatives from eq.(\ref{zcanmf})
with the result,
$$
{ \delta^2 s_C \over \delta \rho({\vec x}) \; \delta \rho({\vec y}) }=
{\delta(\vec x - \vec y) \over \rho({\vec x})} - {\eta \over | \vec x
- \vec y |} \quad , \quad { \delta^2 s_C \over \delta \rho({\vec x})
\; \partial a} = 1
$$
Therefore,
\begin{equation}\label{scuad}
s^{(2)}_C[Y(.),y_0] = \frac12 \; \int  d^3x  \; { Y^2({\vec x}) \over
\rho({\vec x})} - {\eta \over 2} \int  {d^3x  \; d^3y \over |{\vec
x}-{\vec y}|}\;  Y({\vec x})\;  Y({\vec y}) - y_0 \; \int  d^3x \; Y({\vec x})
\end{equation}
Inserting eqs.(\ref{camvar}) and (\ref{s2}) into eq.(\ref{zcanmf}) yields
\begin{equation}\label{gaussy}
e^{\Phi_N(\eta)} \buildrel{ N>>1}\over= e^{-N\,s(\eta)} \; \int\int DY
\, dy_0 \; e^{-N \, s^{(2)}_C[Y(.),y_0] } \; 
\left[ 1 + {\cal O}\left( {1 \over N}\right) \right] 
\end{equation}
where $ s(\eta) \equiv s_C[\rho_s(.),a_s,\eta] $ stands for the value of the
exponent {\bf at} the saddle point. Terms of order higher than
quadratic in $ s_C[\rho(.),a,\eta] $ contribute to the $ 1/N $
corrections.  

The Gaussian functional integral (\ref{gaussy}) can be exactly
computed in terms of the functional determinant of the quadratic form
(\ref{scuad}) [see paper II]. It gives a result of order one ($ N^0 $). 

In the mean field approximation we only keep the dominant order for
large $ N $. Therefore, only the exponent at the saddle point accounts
and  according to eq.(\ref{flib}) we find for the free energy 
\begin{eqnarray}\label{fcampmed}
F &=& F_0 +  N\, T\, s(\eta) + {\cal O}(N^0)\cr \cr
{ p V \over NT} &=& 1 + {\eta \over 3 } { ds \over d \eta} + {\cal O}(N^{-1})
\end{eqnarray}
Hence, in the mean field approximation, the function $ f(\eta) $ is given by
\begin{equation}\label{fetacm}
f_{MF}(\eta)\equiv 1 + {\eta \over 3 } { ds \over d \eta}\; ,
\end{equation}
From eq.(\ref{zcanmf}) we can compute $ s(\eta) $ in terms of the
saddle point solution as follows
\begin{equation}
s(\eta) \equiv s_C[\rho_s(.),a_s,\eta] = -{\eta \over 2} \int  {d^3x  \; d^3y
\over |{\vec x}-{\vec y}|} \;\rho_s({\vec x}) \; \rho_s({\vec y}) + \int
d^3x \; \rho_s({\vec x}) \; \log\rho_s({\vec x}) \; . 
\end{equation}
Using eq.(\ref{eqrho}) we find an equivalent expression that will be useful 
in paper II,
\begin{equation}\label{sdeta}
s(\eta)= {a_s \over 2} +
\frac12 \int  \phi({\vec x}) \; e^{ \phi({\vec x})} \; d^3x \; .
\end{equation}

\subsection{Saddle point evaluation in the microcanonical ensemble}

The extrema of the `effective action' (\ref{accefmc}) dominate the
microcanonical partition function (\ref{ifmcan}) in the large $ N $ limit.
Extremizing eq.(\ref{accefmc}) with respect to $ \rho(.) $ and $ \hat
a $ gives again eqs.(\ref{eqrho}) and (\ref{vincu}), respectively. 

An extra equation follows by extremizing the `effective action'
(\ref{accefmc}) with respect to $ \eta $:
\begin{equation}\label{xieta}
\xi = {3 \over 2 \, \eta_s} - {1 \over2} \int  {d^3x  \; d^3y \over |{\vec
x}-{\vec y}|} \;\rho_s({\vec x}) \; \rho_s({\vec y})
\end{equation}
Going back to dimensionful variables this equation takes the familiar form
$$
E = \frac32 N T - { G \, m^2 \over 2} \int  {d^3q  \; d^3q' \over |{\vec
q}-{\vec q}\,'|} \;N\rho({\vec q}) \; N\rho({\vec q}\,')
$$
That is, eq.(\ref{xieta}) enforces the fixed value of the energy in the
microcanonical ensemble.

Therefore, the stationary point equations in the microcanonical and
canonical ensembles {\bf are identical}. Both ensembles yield the same
results in the $ N \to \infty $ limit in their common region of
validity. We derive the domain of validity of the mean field
approach for each of the three statistical ensembles in paper II. That is, the 
regions where all fluctuations around it decrease its statistical
weight within their common region of validity. 

\bigskip

In order to evaluate the functional integral for the microcanonical
partition function (\ref{ifmcan})
\begin{equation}\label{ifmcan2}
w(\xi,N) \buildrel{ N>>1}\over= \int\int D\rho\; d{\hat a} \;{ d
\eta \over 2 \pi i}\; e^{-Ns_{MC}[\rho(.),\hat a, \eta]}
\end{equation}
we expand the `effective action' $ s_{MC}[\rho(.),\hat a, \eta] $
around the stationary point $ \rho_s(.),{\hat a}_s, \eta_s $ to second
order. This gives
\begin{equation}\label{gausmic}
w(\xi,N) \buildrel{ N>>1}\over=e^{-N\,s(\eta)} \; \int\int DY
\, dy_0 \;{ d {\tilde \eta} \over 2 \pi i}\; e^{-N \,
s^{(2)}_{MC}[Y(.),y_0,{\tilde \eta}] } \;  \left[ 1 + {\cal O}\left( {1 \over
N}\right) \right]  
\end{equation}
where $ Y(.) $ and $ y_0 $ are defined by eq.(\ref{camvar}) and we set $ \eta =
\eta_s + {\tilde \eta} $. The second order piece of the `effective
action' takes now the form
\begin{equation}\label{s2mc}
s^{(2)}_{MC}[Y(.),y_0,{\tilde \eta}] = s^{(2)}_{C}[Y(.),y_0]- {\tilde
\eta} \int  {d^3x  \; d^3y \over |{\vec
x}-{\vec y}|} \;\rho_s({\vec x}) \; Y({\vec y})- {3 \over 4 \,
\eta_s^2}\, {\tilde \eta}^2 \; .
\end{equation}
The Gaussian functional integral in eq.(\ref{gausmic}) yields a
contribution of order one ($N^0$) [see paper II]. The dominant (mean
field) contribution, $ e^{-N\,s(\eta)} $, {\bf exactly coincides} with
the mean field result in the canonical ensemble [eq.(\ref{gaussy})]
Therefore, the canonical and microcanonical ensembles yields {\bf
identical} physical magnitudes and the same equation of state in the
mean field limit. 

\subsection{Spherically symmetric case}

We shall consider here the spherically symmetric case where
eq.(\ref{puntoest}) takes the form
\begin{equation}\label{eqexpr}
{d^2\phi\over dR^2} + \frac2{R} {d \phi\over dR} +  4\pi \eta \;
e^{\phi(R)} = 0\; .
\end{equation}
where we work on an {\bf unit volume} sphere instead of an unit volume
cube as in eq.(\ref{variar}). Therefore, the radial variable runs in
the interval
$$
0 \leq  R \leq  \left({3 \over4\pi}\right)^{1/3}
$$
It is more convenient to introduce a new radial variable
$$
r \equiv R \left({4\pi\over 3}\right)^{1/3}
$$
such that $ 0 \leq  r \leq 1 $.

The saddle point equation (\ref{eqexpr}) takes then the form
\begin{equation}\label{eqexpr2}
{d^2\phi\over dr^2} + \frac2{r} {d \phi\over dr} +  4\pi \eta^R \;
e^{\phi(r)} = 0\; .
\end{equation}
where
\begin{equation}\label{defetaR}
\eta^R \equiv \eta \; \left({4\pi\over 3}\right)^{1/3}  =
1.61199\ldots \; \eta 
\quad \mbox{and} \quad  e^{\phi(r)} = e^{\phi(R)}\; {3 \over4\pi} \; .
\end{equation}

\bigskip

In order to have a regular solution at $ r = 0 $ one has to impose
\begin{equation}\label{fip0}
\phi'(0)=0 \; .
\end{equation}
Otherwise, the second term in eq.(\ref{eqexpr}) diverges for $ r \to 0 $. 

In the spherically symmetric case, the constraint (\ref{vincu2}) becomes 
\begin{equation}\label{vincuR}
\phi'(1)= -\eta^R \; .
\end{equation}

\begin{figure}
\begin{turn}{-90}
\epsfig{file=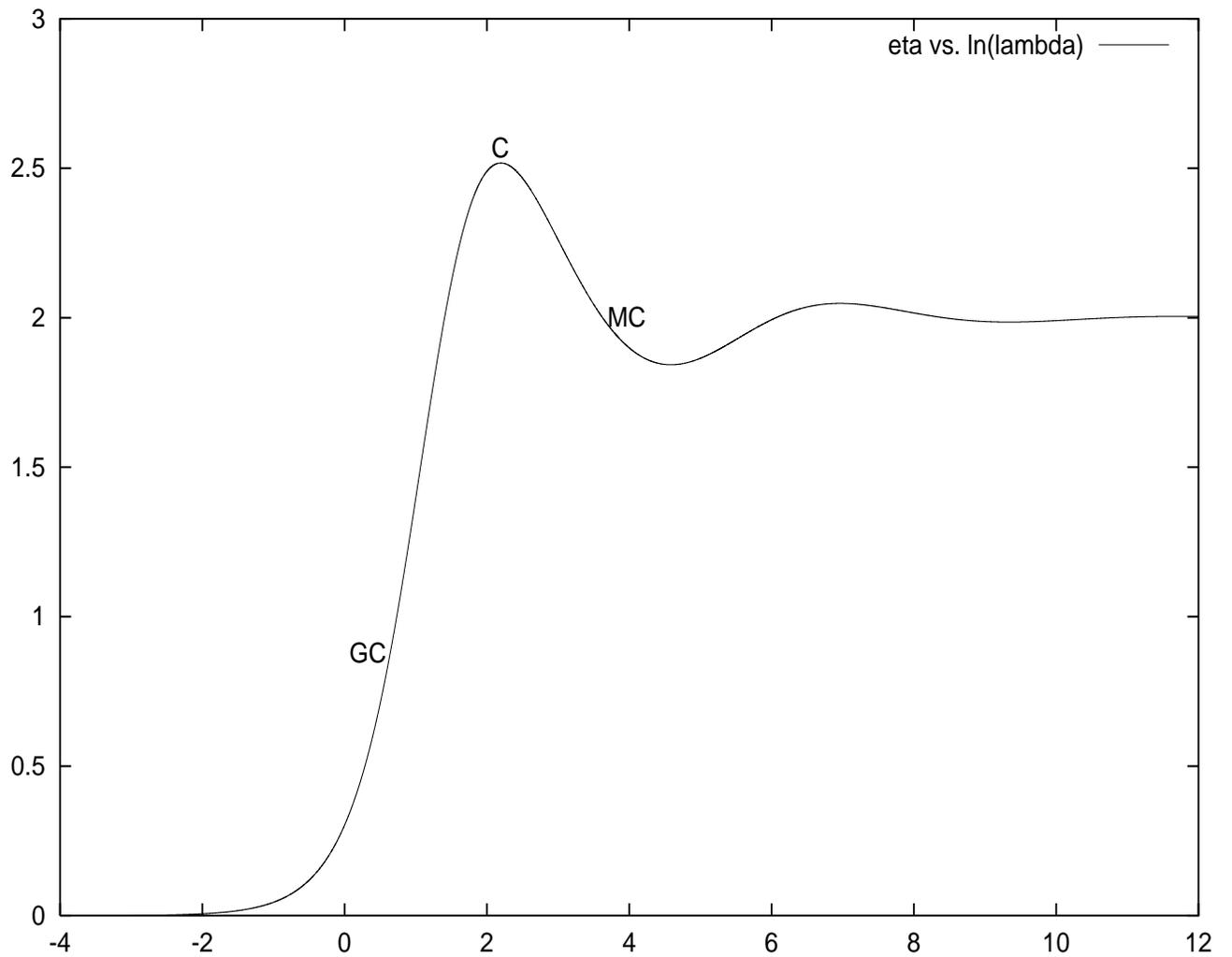,width=14cm,height=18cm} 
\end{turn}
\caption{ $ \eta^R $ as a function of the uniformizing scale  variable
$ \log \lambda $ according to
eq.(\ref{lambaxi}). Notice the maximum of $ \eta^R $ as  $ \eta_C^R =
2.517551\ldots $. The region beyond the point MC, ($\ln \lambda_{MC} =
3.53698\ldots $), is unphysical as we discuss in paper II \label{fig7}} 
\end{figure}

\bigskip

Using the scale covariance  (\ref{coves}) we can express $ \phi(r) $ as 
\begin{equation}\label{fixi}
\phi(r) = \log\left({\lambda^2 \over  4\, \pi \,\eta^R}\right) +
\chi(\lambda \, r)
\end{equation}
where 
\begin{equation}\label{ecuaxi}
\chi''(\lambda) + {2 \over \lambda} \, \chi'(\lambda) +
e^{\chi(\lambda)} = 0 \quad , \quad \chi'(0) = 0 
\end{equation}
This equation is invariant under the transformation:
\begin{equation}\label{invchi}
\lambda \Rightarrow \lambda \; e^{\alpha} \quad \, \quad 
\chi(\lambda)\Rightarrow \chi(\lambda) - 2 \; \alpha \; ,
\end{equation}
where $ \alpha $ is a real number. Hence, we can set 
$ \chi(0) \equiv 0 $ without loosing generality.

\medskip

$ \chi(x) $ is independent of $ \eta^R $, and  $ \lambda $ is related to
 $\eta^R$ through eq.(\ref{vincuR})
\begin{equation}\label{lambaxi}
\lambda \; \chi'(\lambda ) = -\eta^R \; .
\end{equation}
Since $ \lambda $ and $ \eta^R $ are always positive, $ \chi(\lambda)
$ is a monotonically decreasing function of $ \lambda $.

Eq.(\ref{ecuaxi}) can be easily solved for small arguments as
$$
\chi(x) = - {x^2 \over 6}  + {x^4 \over 120} +
{\cal O}(x^6)
$$
Hence, in the dilute limit eq.(\ref{lambaxi}) relating $ \eta^R $ with
$ \lambda $ gives 
\begin{equation}\label{lamchi}
\eta^R = {\lambda^2 \over 3}  -{\lambda^4 \over 30} +  {\cal
O}(\lambda^6) \; .
\end{equation}
For large argument, the solution of eq.(\ref{ecuaxi}) takes the asymptotic
form\cite{chandra} 
\begin{equation}\label{xgde}
\chi(x) = \log{2 \over x^2} + {A \over \sqrt{x}} \cos\left({\sqrt7
\over 2} \log{x} + B\right) \left[ 1 +  {\cal O}\left({1 \over x}\right)\right]
\end{equation}
where $ A $ and $ B $ are numerical constants. Using eq.(\ref{lambaxi}) this
gives for $ \eta^R $  
\begin{equation}\label{asieta}
\eta^R = 2 + {C \over \sqrt{\lambda}} \cos\left({\sqrt7
\over 2} \log{\lambda} + D\right) \left[ 1 +  {\cal O}\left({1 \over
\lambda}\right)\right] 
\end{equation}
where $ C $ and $ D $ are constants related to  $ A $ and $ B $. 
By numerically solving eq.(\ref{ecuaxi}) we find
$$
C = 1.667\ldots \; .
$$
It must be noticed, however, that the mean field solution is unphysical
for $ \lambda>\lambda_{MC} = 34.36361\ldots $ as we shall see in
paper II. Anyway, we see from fig. \ref{fig7} that $ \eta^R $ approaches
very fast its asymptotic behaviour (\ref{asieta}) for $ \log\lambda
> 2 $. 

\begin{figure}
\begin{turn}{-90}
\epsfig{file=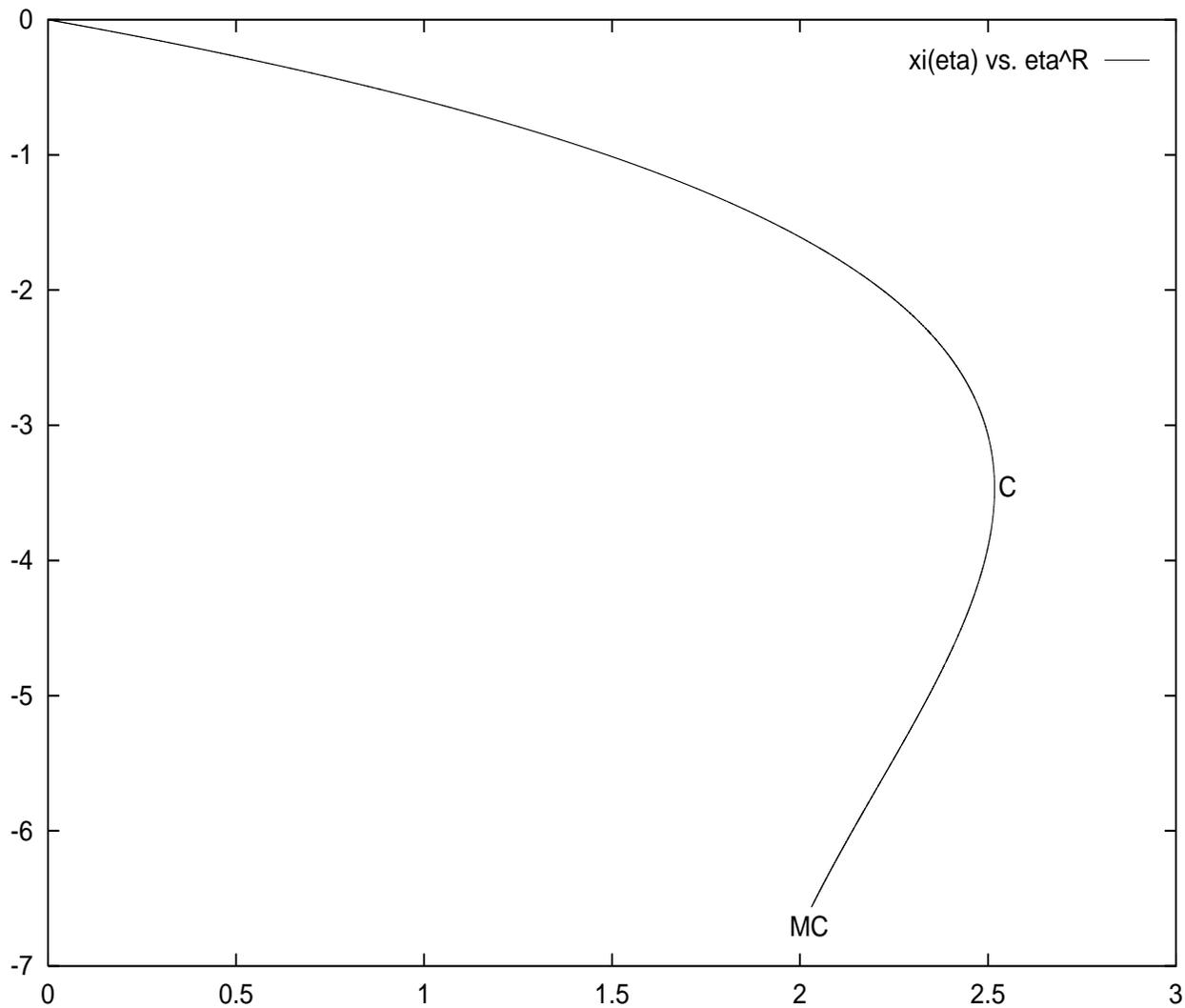,width=14cm,height=18cm} 
\end{turn}
\caption{ $\chi(\lambda(\eta^R)) = \log { p(0) \over p(1) } = \log {
\rho(0) \over \rho(1) } $ as a function of $ \eta^R $. 
\label{fxieta}}
\end{figure}

We plot in fig. \ref{fxieta} $ \chi(\lambda(\eta^R)) $ as a function of
$ \eta^R $. 

\bigskip

In the spherically symmetric case the integral over the angles in
eq.(\ref{eqrho}) is immediate with the result,
\begin{equation}\label{ecuint}
\phi(r) = a_s + 4\pi \eta^R \left[ {1 \over r} \int_0^r {r}'^{\,2}\;  dr' \;
e^{\phi(r')} + \int_r^1 r' \;  dr' \; e^{\phi(r')} \right]
\end{equation}
Deriving with respect to $ r $ yields,
$$
{ d\phi(r) \over dr} = - \frac{4\pi \eta^R}{r^2}\int_0^r {r'}^{\,2}\;  dr'
\;e^{\phi(r')} 
$$
This again shows that $ \phi(r) $ is a monotonically decreasing function of
$ r $ [see above, eq.(\ref{lambaxi})].

Setting $ r = 1 $ in eq.(\ref{ecuint}) leads to the relation
$$
\phi(1) = a_s + 4\pi \eta^R \int_0^1 r^2\;  dr  \; e^{\phi(r)}
$$
Using now the constraint (\ref{vincu}) allows us to compute the Lagrange
multiplier $ a $ at the saddle point
\begin{equation}\label{as}
a_s = \phi(1) - \eta^R
\end{equation}

The particle density in MF is given by
$$
\rho(r) =  e^{\phi(r)} = {\lambda^2 \over  4\, \pi
\,\eta^R}\;e^{\chi(\lambda \, r)} \; , \; 0\leq r \leq 1 \; .
$$
Since $ \chi(\lambda ) $ monotonically decreases with $ \lambda $, 
the particle density monotonically decreases
with $ r $ for fixed $ \eta^R $.

\bigskip

Let us now compute $ s(\eta^R) $ [the exponent in eq.(\ref{zcanmf}) at
the saddle point] for the spherically symmetric case. We find from
eq.(\ref{sdeta}) 
\begin{eqnarray}\label{gorda}
s(\eta^R) &=& \frac12 \left[ \phi(1) - \eta^R \right] + 2 \pi \int_0^1
r^2\;  dr \; \phi(r)\; e^{\phi(r)} \cr \cr
 &=&\log\left({\lambda^2 \over  4\, \pi \,\eta^R}\right) +\chi(\lambda)
- {\eta^R \over 2} + {1 \over 2 \; \lambda \; \eta^R }\int_0^{\lambda} x^2 \;
dx \; [\chi'(x)]^2
\end{eqnarray}
where we integrated by parts and used eqs.(\ref{fixi})-(\ref{lambaxi}).

The integral in the r.h.s. of eq.(\ref{gorda}) can be computed in
closed form [see appendix B] with the result,
$$
s(\eta^R) = \log\left({\lambda^2 \over  4\, \pi \,\eta^R}\right)
+\chi(\lambda) + 3 -\eta^R - {\lambda^2 \over \eta^R}\; e^{\chi(\lambda)}
$$
Inserting now $ s(\eta^R) $ into eq.(\ref{fetacm}) and using
eqs.(\ref{ecuaxi})-(\ref{lambaxi}) yields after calculation
\begin{eqnarray}\label{fetexpl}
f_{MF}(\eta^R) &=& {\lambda^2 \over 3 \, \eta^R}\; e^{\chi(\lambda)}
\quad , \\ \cr
s(\eta^R) & = &  3[1 - f_{MF}(\eta^R)]- \eta^R+ \log\left[ {3
f_{MF}(\eta^R)\over 4\pi}\right] \nonumber
\end{eqnarray}
Notice that $ f_{MF}(\eta^R) $ as well as the other physical
quantities are invariant under the transformation (\ref{invchi}) as it
must be. 

\bigskip
\begin{figure}
\begin{turn}{-90}
\epsfig{file=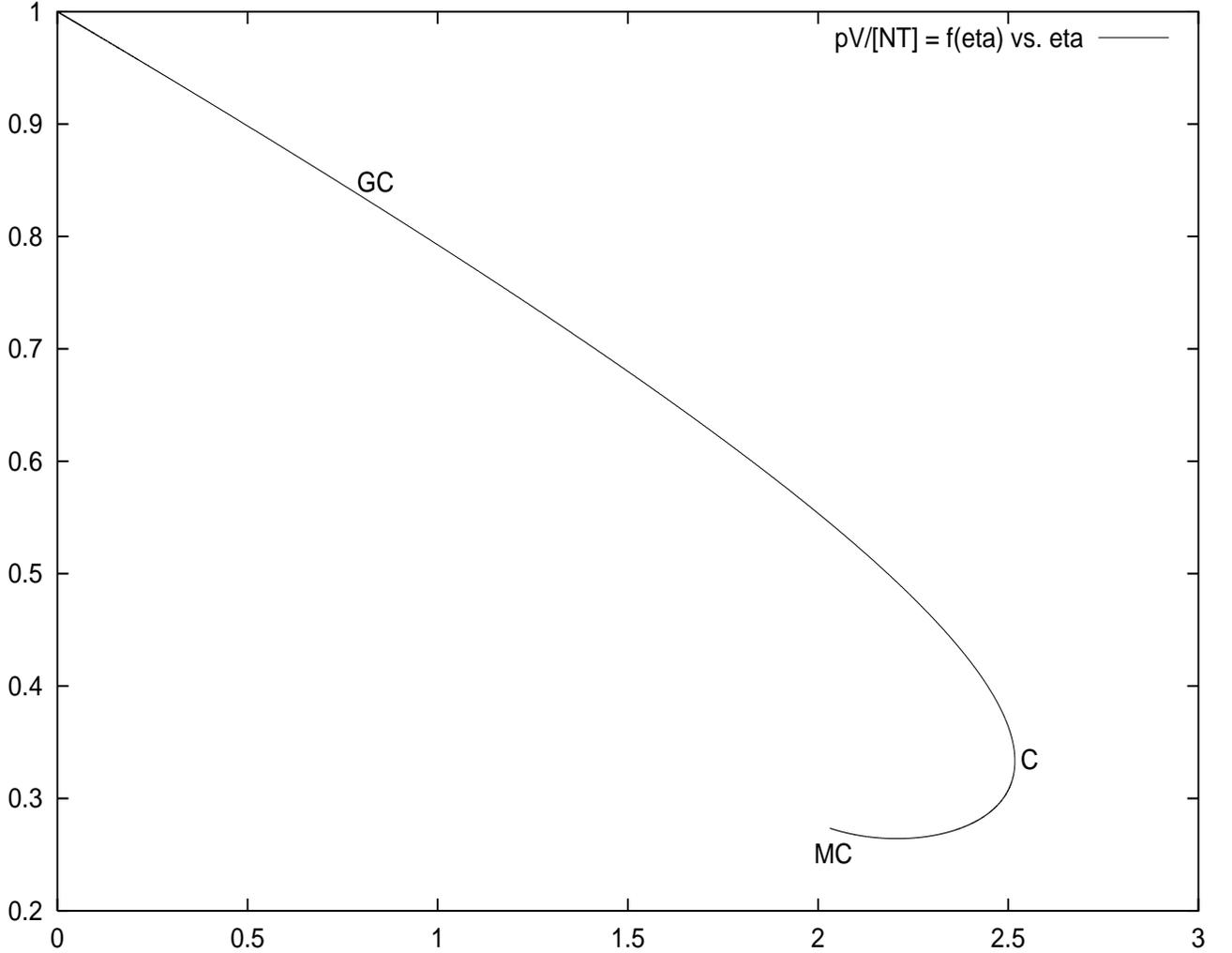,width=14cm,height=18cm} 
\end{turn}
\caption{ $f_{MF}(\eta^R) = P V/[ N T]$ as a function of $\eta^R$ in the MF
approximation [eq.(\ref{abel})].  $f_{MF}(\eta^R)$ has a square root
branch point at $ \eta^R_C $. The points GC, C and MC indicate the
transition to the collapsed phase for each ensemble (grand canonical,
canonical and microcanonical, respectively): $ \eta^R_{GC} =
0.797375\ldots, \; \eta^R_{C} = 2.517551\ldots, \; \eta^R_{MC} =
2.03085\ldots $ (notice that $ \eta^R_{MC} $ is in the second Riemann sheet). 
Since $E/[3NT] = f_{MF}(\eta^R) -\frac12 $, this plot also shows
the energy per particle as a function of $\eta^R$. Furthermore, the
particle density at the surface is proportional to $f_{MF}(\eta^R)$
[eq.(\ref{fide1})]. \label{fig5}} 
\end{figure} 

It follows from eqs.(\ref{ecuaxi}), (\ref{lambaxi}) and (\ref{fetexpl}) that
$ f_{MF}(\eta^R) $ obeys the {\bf first} order non-linear differential
equation 
\begin{equation}\label{abel}
\eta^R(3f_{MF}-1)f'_{MF}(\eta^R)+(3f_{MF}-3+\eta^R) f_{MF} = 0 \;.
\end{equation}
which reduces to an Abel equation of first kind\cite{kam}. 

We thus find that in the mean field approximation all thermodynamic quantities 
follow from the resolution of the {\bf single} first order
non-linear differential eq.(\ref{abel}) with the initial condition
$f_{MF}(0) = 1 $.  

Integrating eq.(\ref{abel}) with respect to $ \eta^R $ yields,
$$
3 \int_0^{\eta^R} {dx \over x} [1 - f_{MF}(x)] = 3 [f_{MF}(\eta^R) - 1
] + \eta^R - \log f_{MF}(\eta^R)
$$
Further useful relations follow from eqs.(\ref{fixi}) and (\ref{fetexpl})
\begin{equation}\label{fide1}
\phi(1) = \log\left[ {3 \, f_{MF}(\eta^R) \over 4 \, \pi }\right]
\quad , \quad \rho(1) = {3 \over 4 \, \pi } \; f_{MF}(\eta^R) \; .
\end{equation}
That is, the particle density at the surface ($r=1$) is proportional
to $ f_{MF}(\eta^R) $.

We can then write the different physical magnitudes in the MF approximation as
\begin{eqnarray}\label{abel2}
{p V \over NT} &=& f_{MF}(\eta^R) \cr \cr
{F - F_0 \over NT}  &=& 3 [1 - f_{MF}(\eta^R)] - \eta^R + \log f_{MF}(\eta^R)
\cr \cr
{S - S_0 \over N} &=& 6 [f_{MF}(\eta^R)-1] + \eta^R - \log f_{MF}(\eta^R)
\\ \cr
{E \over NT} &=& 3 [f_{MF}(\eta^R)-\frac12] \nonumber
\end{eqnarray}
where we used eqs.(\ref{pVnT}), (\ref{enlib}), (\ref{ecan}) and (\ref{entro}).

We derive in appendix C the properties of the function $ f_{MF}(\eta^R)
$ from the differential equation (\ref{abel}). One easily obtains for
small $ \eta^R $ (dilute regime),
$$
f_{MF}(\eta^R) = 1 - \frac{\eta^R}5 - \frac{(\eta^R)^2}{175} + {\cal
O}([\eta^R]^3)\; . 
$$
These terms exactly coincide with the perturbative calculation in the
dilute regime for spherical symmetry [see eq.(\ref{esfera}),
(\ref{petach}) and (\ref{defetaR})].  

We plot in fig. 1 $ f_{MF}(\eta^R) $ as a function of $\eta^R$ obtained
by solving eq.(\ref{abel}) by  the Runge-Kutta method. We see that $
f_{MF}(\eta^R) $ is a {\bf monotonically decreasing} function of $
\eta^R $ for $ 0 < \eta^R < \eta^R_C $. At the point $ \eta^R =
\eta^R_C $, the derivative $ f_{MF}'(\eta^R) $ takes the value $
-\infty $. It then follows from eq.(\ref{abel}) that 
$$
f_{MF}(\eta^R_C) = \frac13 \; .
$$
At the point $ \eta^R_C $ the series expansion for $ f_{MF}(\eta^R) $ in
powers of $  \eta^R $ diverges. Both, from the ratio test on its
coefficients and from the Runge-Kutta solution, we find that
\begin{equation}\label{etacr}
\eta^R_C = 2.517551\ldots 
\end{equation}
From eq.(\ref{abel}) we find that $ f_{MF}(\eta^R) - \frac13 $ has a
square root behaviour around $ \eta^R = \eta^R_C $:
$$
f_{MF}(\eta^R) \buildrel{ \eta^R \uparrow \eta^R_C}\over= \frac13 +
\sqrt{2(\eta^R_C-2) \over 9 \, \eta^R_C} \; \sqrt{\eta^R_C-\eta^R} 
+{2 \, (\eta^R_C-1) \over 7 \, \eta^R_C} \; (\eta^R_C-\eta^R)
+ {\cal O}\left[(\eta^R_C-\eta^R)^{3/2}\right]
$$
Inserting the numerical value (\ref{etacr}) for $ \eta^R_C $ yields,
\begin{equation}\label{cercaetac}
f_{MF}(\eta^R) \buildrel{ \eta^R \uparrow \eta^R_C}\over= \frac13 +
0.213738\ldots \sqrt{\eta^R_C-\eta^R} +0.172225\ldots\; (\eta^R_C-\eta^R)
+ {\cal O}\left[(\eta^R_C-\eta^R)^{3/2}\right]
\end{equation}
We see that $ f_{MF}(\eta^R) $ becomes complex for $ \eta^R > \eta^R_C
$. Recall that in the Monte Carlo simulations the  gas phase collapses at
the point $ \eta^R_T < \eta^R_C $. 

\bigskip

\begin{figure}
\begin{turn}{-90}
\epsfig{file=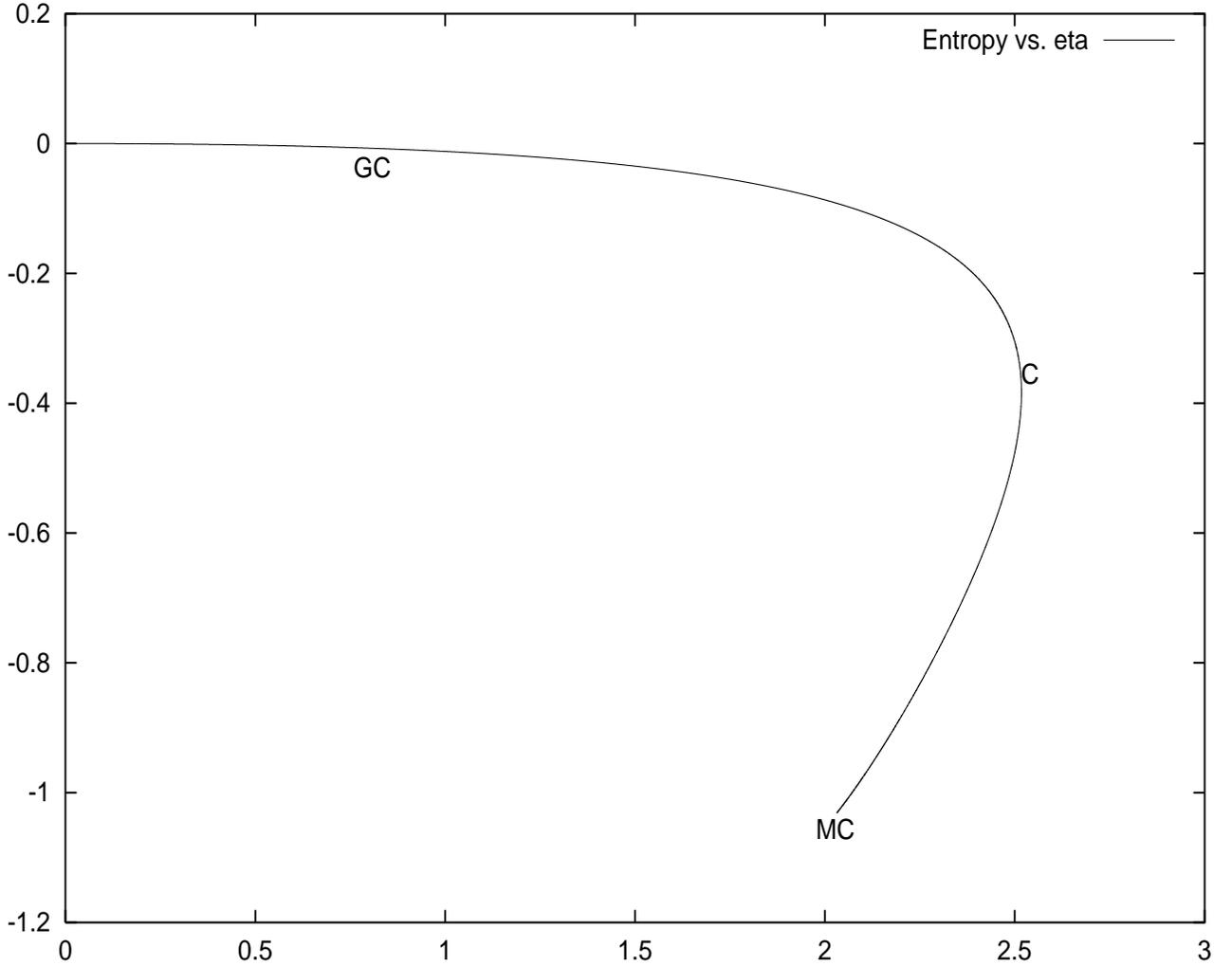,width=14cm,height=18cm} 
\end{turn}
\caption{ The entropy per particle minus the ideal gas value 
 as a function of $\eta^R$ in the MF approximation [eq.(\ref{abel2})].
\label{fig8}} 
\end{figure}

\begin{figure}
\begin{turn}{-90}
\epsfig{file=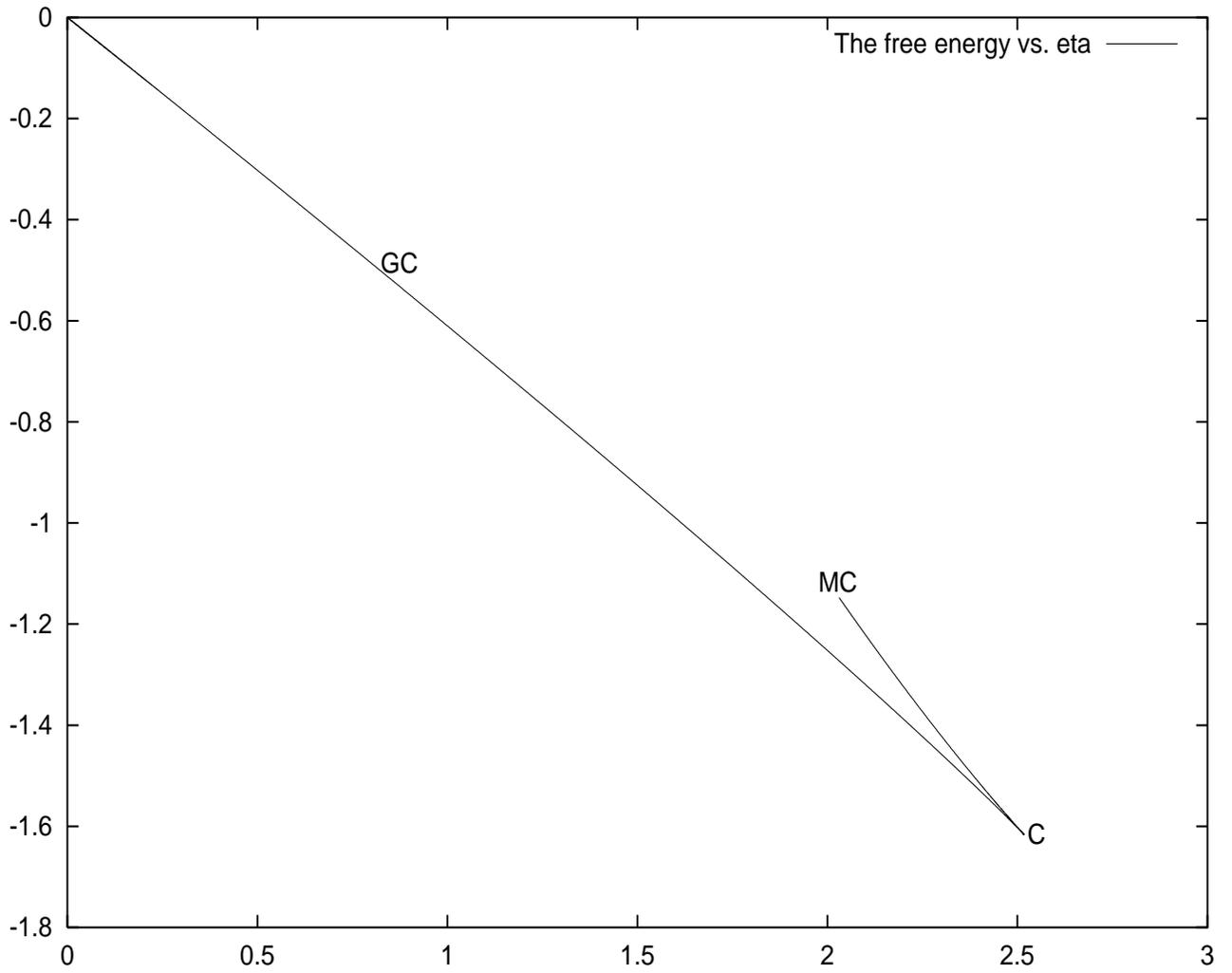,width=14cm,height=18cm} 
\end{turn}
\caption{ $ {F - F_0 \over NT} $ as a function of $ \eta $ in the MF
approximation [eq.(\ref{abel2})]. \label{fig9}} \end{figure}

From eqs.(\ref{abel2}), we plot $ pV/[NT], \; S/N $ and $ {F - F_0
\over NT} $ as a function of $ \eta^R $ in figs. \ref{fig5}, \ref{fig8} and
\ref{fig9},  respectively. 

The points GC, C and MC correspond to the collapse phase
transition in the grand canonical, canonical and microcanonical
ensembles, respectively. Their positions are determined by the
breakdown of the mean field approximation  through the analysis of the
small fluctuations [see paper II].

\bigskip

$ f_{MF}(\eta^R) $ is a multivalued function of $ \eta^R $ as well as
all physical magnitudes [see eq.(\ref{abel2})].

As noticed before, the CE  only describes  the region between the
ideal gas point, $\eta^R = 0$ and $C$ in fig. 1. The MCE goes beyond 
the point $C$ (till the point $MC$) with the physical 
magnitudes described by the second sheet of the square root in
eqs.(\ref{cercaetac}) (minus sign). We have near $C$ between $C$ and $MC$,
\begin{eqnarray}
&& f_{MF}(\eta^R) \buildrel{ \eta^R \uparrow \eta^R_C}\over= \frac13 -
0.213738\ldots \sqrt{\eta^R_C-\eta^R} + 0.172225\ldots\; (\eta^R_C-\eta^R)
+ {\cal O}\; \left[ (\eta^R_C-\eta^R)^{3/2}\right] \nonumber
\end{eqnarray}

\bigskip

The function $ f_{MF}(\eta^R) $ takes its absolute minimum at $ \eta^R
= \eta^R_{min} = 2.20731\ldots $ in the second sheet where  $
f_{MF}(\eta^R_{min}) = 0.264230\ldots$. 

\bigskip

Since $ f_{MF}(\eta^R) < \frac12 $ implies that the total energy is
negative [see eq.(\ref{abel2})], the gas is in a {\it `bounded state'}
for  $ \eta^R $ beyond  $ \eta^R_2 = 2.18348\ldots  $ in the first sheet.

\bigskip

Since $ \chi(\lambda) $ and $ \eta(\lambda) $ are single-valued
functions of $ \lambda , \; f_{MF}(\eta^R(\lambda) )$ defined by
eq.(\ref{fetexpl}) is also a single-valued function of $ \lambda
$. That is, $ \lambda $ is the {\bf uniformization } variable. All
physical magnitudes are single-valued functions of $ \lambda $. 
On the other hand,  $ \lambda $ is an infinite-valued function of $
\eta^R $ as one sees from fig. \ref{fig7} and eq.(\ref{asieta}).
That is,  $f_{MF}(\eta^R)$ has an infinite number of Riemann
sheets. However, only the first two sheets are physically
realized. The rest are unphysical. A plot of $f_{MF}(\eta^R)$
including all sheets produces a nice spiral\cite{chandra}  converging towards $
\eta^R = 2 \; , f_{MF}(\eta^R) = 1/3 $ for $ \lambda = \infty $ as
follows from eqs. (\ref{xgde}), (\ref{asieta}) and (\ref{fetexpl}). 

\bigskip

$ \lambda $ induces a scale transformation in coordinate space as we
see in eq.(\ref{fixi}) whereas $ \eta^R $ plays the coupling constant
[Recall that $ \eta^R $ is proportional to Newton's gravitational constant]. 

The variation of $ \eta^R $ with respect to $ \lambda $ yields the
renormalization group equation
$$
\lambda {d \eta^R \over d \lambda} = \eta^R \; [ 3 \, f_{MF}(\eta^R) - 1 ]
$$
where we used eqs.(\ref{ecuaxi}), (\ref{lambaxi}) and (\ref{fetexpl}).
Here $ \eta^R \; [ 3 \, f_{MF}(\eta^R) - 1 ] $ plays the role of the
renormalization group beta function. We see that it has two fixed
points at $ \eta^R = 0 $ and at $ \eta^R = \eta^R_C $. [See
fig. \ref{fig7} where the running of  $ \eta^R $ with $ \lambda $ is
exhibited].  

We find from eqs.(\ref{lamchi}) and (\ref{cercaetac})
near these fixed points
$$
\eta^R \buildrel{ \lambda \to 0}\over= {\lambda^2 \over 3}
$$
$$
\eta^R\buildrel{ \lambda \to \lambda_C}\over= \eta^R_C - { \eta^R_C
(\eta^R_C - 2) \over 2 \;  \lambda_C^2 }\; ( \lambda - \lambda_C)^2
$$
where the coefficient has the numerical value $ { \eta^R_C
(\eta^R_C - 2) \over 2 \;  \lambda_C^2 } = 0.0085515\ldots $.

\begin{figure}[t] 
\begin{turn}{-90}
\epsfig{file=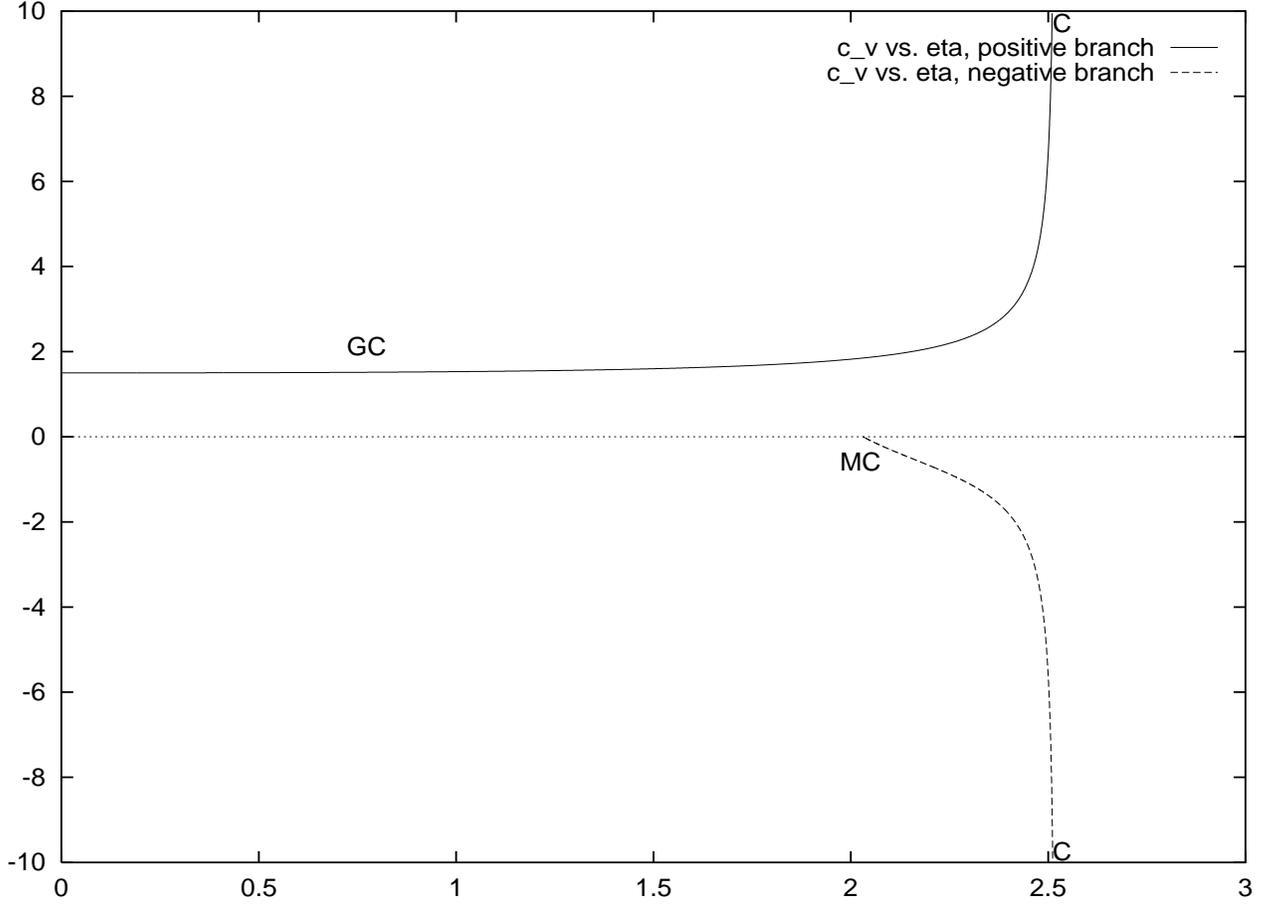,width=12cm,height=18cm} 
\end{turn}
\caption{ $ (c_V)_{MF} $ as a function of $ \eta^R $ from mean field
eq.(\ref{cvmf}). Notice that $ (c_V)_{MF} $ diverges at the point C, that is
for $ \eta^R_C = 2.517551\ldots $ \label{fig1}} 
\end{figure}

\subsection{Canonical vs. Grand Canonical Ensembles in the Mean Field
Approximation} 

We have seen that the stationary point equations and their respective
solutions are closely related in the canonical and grand canonical
ensembles [eqs.(\ref{ecdelgc})-(\ref{intsgc}) and
(\ref{vincu2})-(\ref{cangcan})]. 

Let us now show that physical quantities obtained from both ensembles do
coincide in the mean field approximation.

From eq.(\ref{defKC}) and (\ref{cangcan}) we find that
\begin{equation}\label{calK}
K(\eta)= \int  \phi({\vec x}) \; e^{ \phi({\vec x})} \; d^3x + \log
C(\eta) \; .
\end{equation}
[Recall that $\int  e^{ \phi({\vec x})} \; d^3x = 1 $].

In the spherically symmetric case this integral takes the form
\begin{equation}\label{esfKC}
4 \pi \int_0^1 r^2\;  dr \; \phi(r)\; e^{\phi(r)} = \phi(1) + {1 \over
\eta^R } \int _0^1 r^2\;  dr \; \left({d \phi \over dr}\right)^2 = 
6 \,[1 - f_{MF}(\eta^R)] - \eta^R + \log\left[ {3 f_{MF}(\eta^R)\over
4\pi}\right] 
\end{equation}
where we integrated by parts and used eqs.(\ref{fixi}),
(\ref{lambaxi}) and appendix B. 

From eqs.(\ref{calK}) and (\ref{esfKC}) we  find
$$
K(\eta^R)- \log C(\eta^R) = 6 \,[1 - f_{MF}(\eta^R)] - \eta^R +
\log\left[ {3 f_{MF}(\eta^R)\over 4\pi}\right] \; .
$$
Inserting this result into the linear differential equations (\ref{ecKC}) leads
to the solution,
\begin{equation}\label{solKC}
C(\eta^R) = {4 \, \pi \over 3 }\; { \exp[\eta^R] \over
f_{MF}(\eta^R)} \quad \mbox{and}  \quad K(\eta^R)=6\,[1 - f_{MF}(\eta^R)]
\end{equation}
We then find from eqs.(\ref{cangcan}), (\ref{as}) and (\ref{fide1}) that
\begin{equation}\label{Fisfi}
\log C(\eta) = -a_s \; .
\end{equation}
Combining eq.(\ref{solKC}) with eqs.(\ref{ugc}),
(\ref{efegc})-(\ref{entgc}) and (\ref{prgc}) shows that the canonical
and the grand canonical ensembles yields {\bf identical} physical magnitudes
(pressure, energy, entropy, free energy, specific heats,
compressibilities, speed of sound) and the same equation of state in
the mean field approximation. 

The thermodynamical potential\cite{llms},
$$
\Omega \equiv -T \, \log{\cal Z}_{GC} = N \,[3\, f_{MF}(\eta^R)- 2]
$$
is {\bf not} equal to $ - PV $. That is, here $ \Omega \neq - PV $
and we have instead
$$
\Omega + PV = 2\, NT [1 - f_{MF}(\eta^R)]
$$
This relation is analogous to eq.(\ref{gibbs2}). $ \Omega $ differs
here from $ - PV $ since for the self-gravitating gas
we have $ N \sim L $  instead of the usual relation $ N \sim L^3 $. 

\section{Specific Heats, Speed of Sound and Compressibility}

The specific heat at constant volume in the mean field approximation takes
the form
\begin{equation}\label{cvmf}
(c_V)_{MF}= 6 \, f_{MF}(\eta^R)- \frac72 + \eta^R + {\eta^R - 2 \over
3 \, \, f_{MF}(\eta^R) - 1}
\end{equation}
where we used eqs.(\ref{ceV}) and (\ref{abel}). 

\begin{figure}[t] 
\begin{turn}{-90}
\epsfig{file=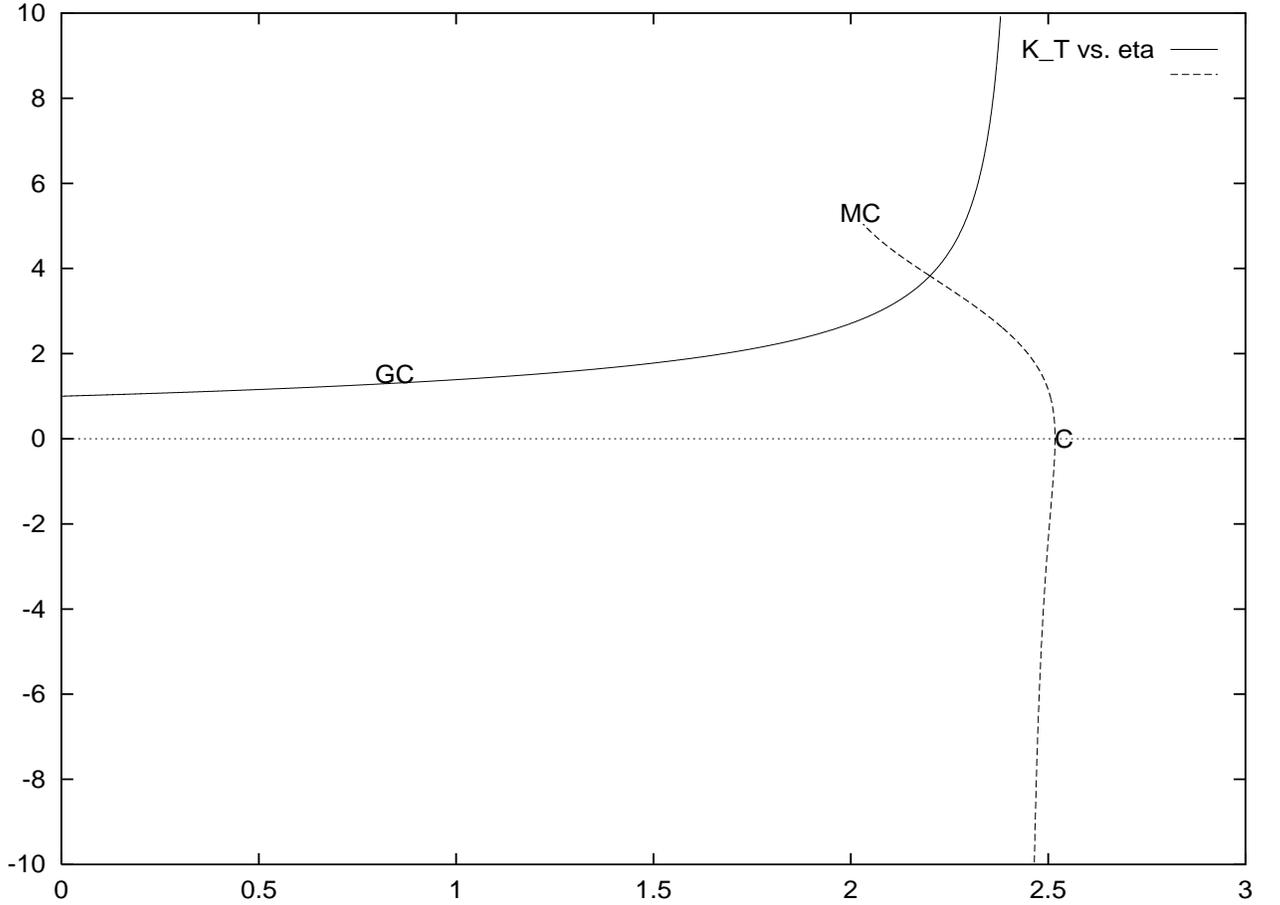,width=12cm,height=18cm} 
\end{turn}
\caption{ $ (\kappa_T)_{MF} $ as a function of $ \eta $ from mean field
eq.(\ref{ktmf}). Notice that $ (\kappa_T)_{MF} $ diverges at $ \eta^R
= \eta_0^R = 2.43450\ldots $ \label{fig11}}  
\end{figure}

\begin{figure}[t] 
\begin{turn}{-90}
\epsfig{file=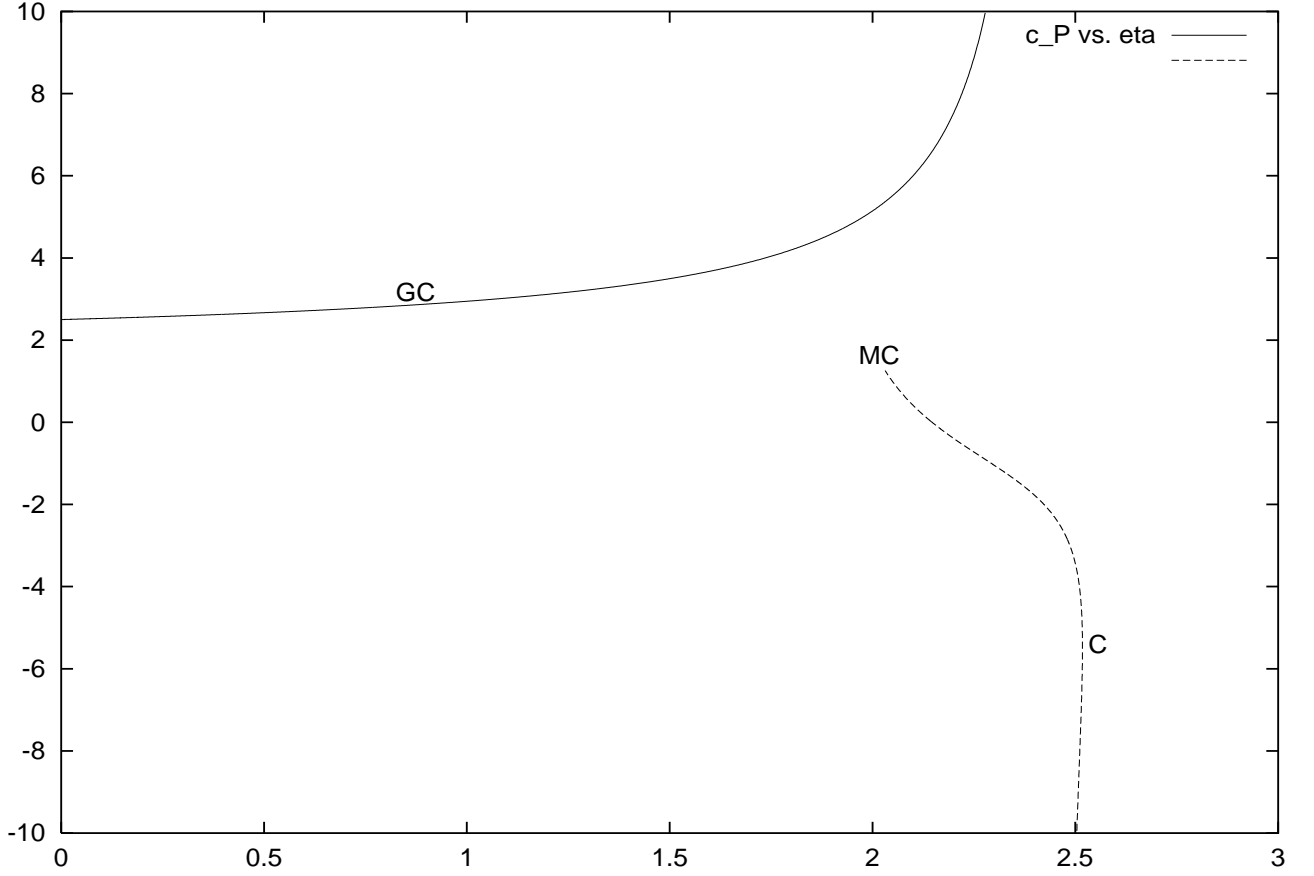,width=12cm,height=18cm} 
\end{turn}
\caption{ $ c_P $ as a function of $ \eta $ from mean field
eq.(\ref{cpmf}). Notice 
that $ c_P $ diverges at $ \eta^R = \eta^R_0 = 2.43450\ldots $ \label{fig10}} 
\end{figure}

We plot in Fig. \ref{fig1} eq.(\ref{cvmf}) for $ (c_V)_{MF} $ as a function of
$ \eta $. We see that $ (c_V)_{MF} $ increases with $ \eta $ till it tends to
$ + \infty $ for  $ \eta^R \uparrow \eta^R_C $. It has a square-root
branch point at the point C. In the stretch C-MC (only  physically
realized in the 
microcanonical ensemble), $ (c_V)_{MF} $ becomes negative. We shall
not discuss here the peculiar properties of systems with negative $
C_V $ as they can be find in refs.\cite{lynbell,lynbell2,bt}

From eqs.(\ref{cercaetac}) and (\ref{cvmf}) we obtain the following behaviour
near the point $C$ in the positive (first) branch
\begin{eqnarray}\label{mfcri}
(c_V)_{MF}&\buildrel{ \eta^R \uparrow \eta^R_C}\over=& 0.80714\ldots
(\eta^R_C-\eta^R)^{-1/2} - 0.19924\ldots+ {\cal O}(\sqrt{\eta^R_C-\eta^R})
\end{eqnarray}
and between $C$ and $MC$ in the negative (second) branch 
\begin{eqnarray}
(c_V)_{MF}&\buildrel{ \eta^R \uparrow \eta^R_C}\over=&
-0.80714\ldots(\eta^R_C-\eta^R)^{-1/2} - 0.19924\ldots+{\cal
O}(\sqrt{\eta^R_C-\eta^R})\nonumber 
\end{eqnarray}
Finally,  $ (c_V)_{MF} $ vanishes at the point MC $ \eta^R_{MC}=
2.03085\ldots $. 
 
\begin{figure}[t] 
\begin{turn}{-90}
\epsfig{file=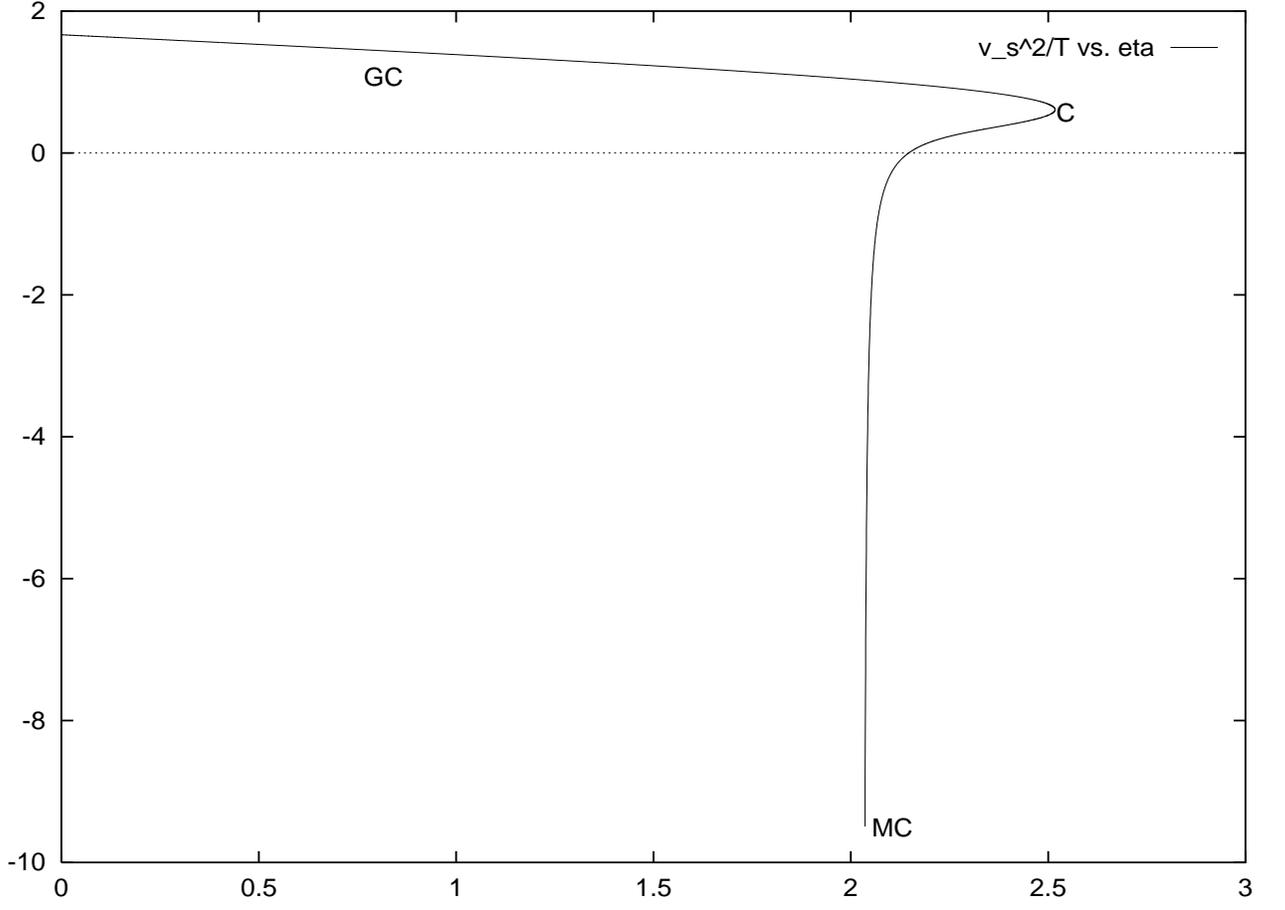,width=12cm,height=18cm} 
\end{turn}
\caption{The speed of sound squared at the surface divided by the temperature,
$ v_s^2 / T $,  as a function of $ \eta $ from mean field [eq.(\ref{cpmf})].
Notice that $ v_s^2 / T $ takes the value $ 11/18 $ at the critical
point $ \eta=\eta_C $ and becomes negative beyond $ \eta^R_1 =
2.14675\ldots $ in the second sheet.
\label{fig3}} 
\end{figure}

\bigskip

The isothermal compressibility in mean field follows from
eqs.(\ref{KT}) and (\ref{abel})
\begin{equation}\label{ktmf}
(\kappa_T)_{MF} = {3 \over 2  f_{MF}(\eta^R)} \, \left[ 1 + {
\eta^R - 2 \over 6  f_{MF}(\eta^R) - \eta^R}\right]\; .
\end{equation}
We plot $ (\kappa_T)_{MF} $ in fig. \ref{fig11}. We see that
$ (\kappa_T)_{MF} $ is positive for $ 0 \leq \eta^R < \eta^R_0 =
 2.43450\ldots$ where $ (\kappa_T)_{MF} $ diverges. The point $ \eta^R_0 $ 
is defined by the equation
\begin{equation}\label{defeta0}
6  f_{MF}(\eta^R_0) - \eta^R_0 = 0\; .
\end{equation}
We find from eqs.(\ref{abel}) and (\ref{defeta0}) that
\begin{equation}\label{fpeta0}
 f_{MF}'(\eta^R_0) = - \frac12
\end{equation}
$ (\kappa_T)_{MF} $ diverges for $ \eta^R \simeq \eta^R_0 $ as
$$
(\kappa_T)_{MF} \buildrel{\eta^R \simeq \eta^R_0 }\over=
{9 \, (\eta^R_0-2) \over 4 \, \eta^R_0 (\eta^R_0 - \eta^R)} + {\cal O}(1) = {
0.40157\ldots  \over \eta^R_0 - \eta^R} + {\cal O}(1) \; .
$$
$ (\kappa_T)_{MF} $ is  negative for $ \eta^R_0 < \eta^R < \eta^R_C $
and exactly  vanishes at the point $ C $. $(\kappa_T)_{MF}$ then
becomes positive in the stretch between $C$ and $MC$ only  physically
realized in the microcanonical ensemble.   

Notice that the singularity of  $(\kappa_T)_{MF}$ at  $ \eta^R =
\eta^R_0 =  2.43450\ldots $ is before but near the  point $C$. It 
appears as a preliminary signal of the phase transition at
$C$. $\eta^R_0$ is probably the transition point $ \eta_T $ seen with the Monte
Carlo simulations (see fig. 1). (Recall that $  \eta_T \sim 1.515 $
corresponds to  $ \eta^R_T \sim 2.44 $).

\bigskip

It is easy to understand the meaning of a large compressibility. From the
definition (\ref{KT})
\begin{equation}\label{compre}
{\delta V \over V} = - K_T \; \delta p = - \kappa_T \; {V \; \delta p
\over N \, T} \; .
\end{equation}
A large compressibility implies that a small increase in the pressure
($ \delta p \ll NT/V $) produces a large change in the
density of the gas. That means a very soft fluid. 

For {\bf negative} compressibility, eq.(\ref{compre}) tells us that the gas
{\bf increases} its volume when the external pressure on it
increases. This is clearly an unusual behaviour that leads to
instabilities as we shall see below. 

\bigskip

The specific heat at constant pressure in the mean field approximation takes
the form
\begin{equation}\label{cpmf}
(c_P)_{MF}= 12 \, f_{MF}(\eta^R)- \frac32 + {24\left(\eta^R -
2\right) f_{MF}(\eta^R) \over 6 \, \, f_{MF}(\eta^R) - \eta^R}
\end{equation}
where we used eqs.(\ref{ceP}) and (\ref{abel}). We plot  $ (c_P)_{MF}
$ in fig. \ref{fig10}. We see that $ (c_P)_{MF} $ is positive and grows
with $ \eta^R $ till it diverges at the same point where  $ (\kappa_T)_{MF}
$ diverges $ \eta^R = \eta^R_0 = 2.43450\ldots $. That is,
$$
(c_P)_{MF} \buildrel{\eta^R \simeq \eta^R_0 }\over= 
{\eta^R_0 (\eta^R_0-2) \over \eta^R_0 - \eta^R } + {\cal O}(1) = {
1.05779\ldots  \over \eta^R_0 - \eta^R} + {\cal O}(1) \; .
$$
$ (c_P)_{MF} $
becomes negative for $ \eta^R_0 < \eta^R < \eta^R_C $. It keeps
negative in the C-MC section till the point $  \eta^R=\eta^R_1 =
2.14675\ldots $ where it becomes positive. The point $ \eta^R_1 $ is
defined by the equation
\begin{equation}\label{eta1}
24 \, f_{MF}^2(\eta^R_1)+ \left( 4 \, \eta^R_1
-19\right)f_{MF}(\eta^R_1) + \frac{\eta^R_1}2 = 0 \; .
\end{equation}

The speed of sound squared at the surface in the mean field approximation
takes the form
\begin{equation}\label{vsmf}
{v_s^2 \over T} = {  f_{MF}(\eta^R) \over 3 }\, \left[ 4 + { 3\,
f_{MF}(\eta^R) + \frac{\eta^R}2 - 2 \over 6 \, f_{MF}^2(\eta^R)+
\left(\eta^R - \frac{11}2 \right) f_{MF}(\eta^R) + \frac12 } \right]  \;,
\end{equation}
where we used eqs.(\ref{vson}) and (\ref{abel}). We plot 
$ {v_s^2 \over T} $ as a function of $ \eta^R $ 
in fig. \ref{fig3}. We see that  $ {v_s^2 \over T}(\eta^R) $ is
positive and 
decreasing with $ \eta^R $ in the whole interval between $ \eta^R = 0 $
and $C$. At the  point $C$ it takes the value ${v_s^2 \over T}(\eta^R_C) =
11/18 $. Then, 
$ {v_s^2 \over T}(\eta^R)$ decreases between $C$ and $MC$ becoming 
negative at $ \eta^R_1 = 2.14675...$ in the second sheet where it
vanishes. Notice that 
 $ {v_s^2 \over T}(\eta^R)$ and $ (c_P)_{MF} $ vanish at the same
point $ \eta^R_1 $ defined by eq.(\ref{eta1}).

$v_s^2 < 0$ indicates an instability. That is, small density
fluctuations grow exponentially in time instead of propagating
harmonically. It is remarkable that $v_s^2$ becomes negative at $
\eta^R_1 = 2.14675...$  in the second sheet {\bf before but near} the
$MC$ critical point 
$\eta^R_{MC} = 2.03085...$ in the second sheet. Somehow, the change of
sign in $v_s^2$  announces the  $MC$ critical point.

$ {v_s^2 \over T}(\eta^R)$ tends to $ -\infty $ for $ \eta^R \downarrow
\eta^R_{MC} $. Notice that the denominator in eq.(\ref{vsmf}) exactly
vanishes at $ \eta^R = \eta^R_{MC} $ [see Table 1].

\bigskip

The adiabatic compressibility $ \kappa_S $ is not here an independent
quantity. We find from eqs.(\ref{kapa}), (\ref{cvmf}), (\ref{ktmf})
and (\ref{cpmf}),
$$
\kappa_S = { c_V \over c_P} \; \kappa_T = {3 \over  f_{MF}(\eta^R)} \;
{ 12 \,  f_{MF}^2(\eta^R) +(2 \,\eta^R -11)\,  f_{MF}(\eta^R) + 1 \over 48
\,  f_{MF}^2(\eta^R) + (8\, \eta^R -38)\,  f_{MF}(\eta^R) +\eta^R }
$$
That is,
$$
\kappa_S = { T \over v_s^2} \; .
$$

\vspace{1cm}
\begin{turn}{90}
\begin{tabular}{|l|l|l|l|l|l|}\hline
$$ & $$ & $$ & $$ & $$ & \\
 POINT  & $\hspace{0.8cm}\lambda$ & $\hspace{0.8cm}\eta^R$ &
\hspace{2cm} Defining Equation & $ \hspace{0.2cm} f_{MF}(\eta^R) $ &
PHYSICAL MEANING \\ 
$$ & $$ & $$ & $$ & $$ &\\ \hline 
$$ & $$ & $$ & $$ & $$ &\\
$\;$ GC  &   $ 1.7772\ldots $ & $0.797375\ldots $ & $\hspace{1.75cm} 2 - 3\;
\eta^R_{GC}\; f_{MF}(\eta^R_{GC}) = 0 $ & $ 0.836076\ldots $ & Collapse
in the GCE. \\ 
$$ & $$ & $$ & $$ & $$ & $$ \\ \hline 
$$ & $$ & $$ & $$ & $$ &  Energy density \\ 
$\; \; \; \; 3 $ & $ 3.38626\ldots $ & $
1.73745\ldots     $ & $\hspace{1.75cm} 3 - \eta^R + \chi(\lambda) = 0 
$ & $ 0.622424\ldots  $ & vanishes at $r=0$. \\
$$ & $$ & $$ & $$ & $$ & $$ \\ \hline 
$$ & $$ & $$ & $$ & $$ & $$ \\
$\; \; \; \;  2$ & $4.73739\ldots $ & $ 2.18348\ldots$ &\hspace{2.1cm} $2\;
f_{MF}(\eta^R_2) - 1 = 0 $ & $\hspace{0.5cm} 0.5$ & Total Energy vanishes.\\ 
$$ & $$ & $$ & $$ & $$ &$$ \\ \hline
$$ & $$ & $$ & $$ & $$ &$$ \\
$ \; \; \; \; 0$ & $6.45077\ldots  $ & $ 2.43450\ldots$ &\hspace{2.1cm} $6\;
f_{MF}(\eta^R_0) - \eta^R_0 = 0 $ & $0.40575\ldots$ & $\kappa_T$ and
$c_P$ diverge. \\ 
$$ & $$ & $$ & $$ & $$ & $$ \\ \hline 
$$ & $$ & $$ & $$ & $$ & Collapse in the CE. \\
$ \; \; \;  $ C &  $ 8.993195\ldots  $ &  $ 2.517551\ldots  $ &
$\hspace{2.1cm} 1 
- 3 \; f_{MF}(\eta^R_{C}) = 0 $ & $ \hspace{0.5cm} 1/3 $ & 
$ c_V $ diverges. \\ 
$$ & $$ & $$ & $$ & $$ &$$ \\\hline
$$ & $$ & $$ & $$ & $$ & Minimum of\\
$ \; \; $ Min & $22.5442\ldots $ & $2.20731\ldots$ &\hspace{2.1cm} $
f'_{MF}(\eta^R_{min}) = 0 $ &  $ 0.264230\ldots$ & pV/[NT] \\ 
$$ & $$ & $$ & $$ & $$ & in the gas phase\\ \hline
$$ & $$ & $$ & $$ & $$ & $$\\
$ \; \; \; \; 1 $ & $25.7991\ldots $ & $ 2.14675\ldots $ & $ \; \;
 48f_{MF}^2(\eta^R_1) - (38 - 8 \eta^R_1)\, f_{MF}(\eta^R_1) +
$ & $0.265290\ldots$ & $v_s^2$ and $c_P$ vanish.
\\ $$ & $$ & $$ &\hspace{3cm} $  + \eta^R_1 = 0  $ & $$ &\\ \hline 
$$ & $$ & $$ & $$ & $$ & Collapse in the MCE. \\
 $ \; \; $ MC  & $ 34.36361\ldots $ & $ 2.03085\ldots $ &
$ 12f_{MF}^2(\eta^R_{MC}) - (11 - 2 \eta^R_{MC})\, f_{MF}(\eta^R_{MC})+
$  & $ 0.273512\ldots $ & $c_V$ vanishes.
\\ $$ & $$ & $$ & \hspace{3cm} $+ 1 = 0$ & $$ & $$ \\ \hline
\end{tabular}
\end{turn}
\vspace{0.5cm}

{TABLE 1. Values of the critical points in the three ensembles GC, C
and MC (using mean field) and further
characteristic points for spherical symmetry. $ pV/[NT], E $ and $ S \to
-\infty $ for $  \eta^R \uparrow \eta^R_{GC} $ and $ \eta^R \uparrow
\eta^R_C $. Notice that $ \eta_{Min}, \; 
\eta_1 $ and $ \eta_{MC} $ are in the second Riemann sheet.}

\section{Discussion}

We have presented here a set of new results for the self-gravitating thermal
gas obtained by Monte Carlo and analytic methods. They provide a
complete picture for the thermal self-gravitating gas.

Contrary to the usual hydrostatic treatments \cite{chandra,sas}, we {\bf do not
assume} here an equation of state but we {\bf obtain} the equation of state 
from the partition function [see eq.(\ref{pVnT})]. We find at the same time
that the relevant   variable is here $ \eta^R = G m^2 N/[V^{1/3}  T] $.
The relevance of the ratio $ G m^2 /[V^{1/3}  T] $ has been noticed on
dimensionality  grounds \cite{sas}. However, dimensionality arguments alone
cannot single out the crucial factor $ N $ in the variable $ \eta^R $.

The crucial point is that the thermodynamic limit exist if we
let  $ N \to \infty $ and $ V \to \infty $ {\bf keeping $ \eta^R $ fixed}. 
Notice that $ \eta $ 
contains the ratio $ N \; V^{-1/3} $ and not $ N / V $. This means that
in this thermodynamic limit $ V $ grows as $ N^3 $ and thus the volume density
$ \rho = N/ V $ decreases as $ \sim N^{-2} $. $ \eta $ is to be kept fixed
for a  thermodynamic limit to exist in the same way as the temperature.
 $ p V $, the energy $E$, the free energy, the entropy are functions of 
$ \eta $ and $ T $ times $N$. The chemical potential,
specific heat, etc. are just functions of $ \eta $ and $ T $.

We find collapse phase transitions both in the canonical and in the
microcanonical ensembles. They take place at different values of the
thermodynamic variables and are of different nature. In the CE the
pressure becomes large and negative in the collapsed phase.
The  phase transition in the MCE is
sometimes called `gravothermal catastrophe'. We find that the
temperature and pressure increase discontinuously at the MCE transition.
Both are zeroth order phase transitions (the Gibbs free energy is
discontinuous). The two phases cannot coexist in equilibrium since
the pressure has different values at each phase. 

The parameter $ \eta^R $ [introduced in eq.(\ref{defeta})] can be
related to the Jeans length of the system  
\begin{equation}\label{longJ}
d_J = \sqrt{3T \over m} { 1 \over \sqrt{G \, m \, \rho}} \; ,
\end{equation}
where $ \rho \equiv N/V $ stands for the number volume density. Combining
eqs.(\ref{defeta}) and (\ref{longJ}) yields
$$
\eta^R = 3 \left(L \over d_J \right)^2 \; .
$$
We see that the  phase transition  in the canonical ensemble takes
place for $ d_J \sim L $. [The precise numerical value of the
proportionality coefficient depends on the geometry]. For $  d_J > L $
we find the gaseous phase and for $  d_J < L $ the system condenses as
expected. Hence, the collapse phase transition in the canonical
ensemble is related to the Jeans instability.  

The latent heat of the transition ($q$) is {\bf negative} in the CE
transition indicating that the gas releases heat when it collapses
[see eq.(\ref{qsobreT})]. The MCE transition exhibits an opposite
behaviour. The Gibbs free energy increases at the MCE collapse phase
transition (point MC in fig.\ref{fig14}) whereas it decreases at the CE
transition [point T in fig. \ref{fig14}, see eq.(\ref{deltaG})].  Also, the
average distance 
between particles increases at the MCE phase transition whereas it
decreases dramatically in the CE  phase transition. These differences
are related to the MCE constraint keeping the energy fixed whereas in
the CE the system exchanges energy with an external  heat bath keeping
fixed its temperature. The constant energy  constraint in the MCE
keeps the gas stable in a wider domain and makes the collapse
transition softer than in the CE. Notice that the core is much tighter
and the halo much smaller  in the CE than in the MCE [see
figs. \ref{colmc} and  \ref{colc}]. 

\section{Acknowledgements}

One of us (H J de V) thanks M. Picco for useful discussions on Monte
Carlo methods. We thank S. Bouquet for useful discussions and J. Katz
for calling our attention on ref.\cite{katz}. 

\appendix

\section{Functional integration Measure in the Mean Field Approach}

We follow the derivation of ref.\cite{lipa} for the functional
integral measure. We want to recast
\begin{equation}\label{fupa}
e^{\Phi_N(\eta)} = \int_0^1\ldots \int_0^1
\prod_{l=1}^N d^3r_l\;\; e^{ \eta \; u({\vec r}_1,\ldots,{\vec r}_N)}\; ,
\end{equation}

as a functional integral in the large $ N $ limit.

We start by dividing the domain of integration (of unit volume) into $
M $ cells. Each cell is of volume $ \omega_r $ and contains $ k_r $
particles with $ 1 \leq r \leq M $. Therefore,
$$
\sum_{r=1}^M k_r = N \quad , \quad \sum_{r=1}^M \omega_r = 1 \; .
$$
We can thus rewrite the multiple integral (\ref{fupa}) as follows:
$$
e^{\Phi_N(\eta)} = \sum_{k_1, \ldots , k_M} \delta\left(N -
\sum_{r=1}^M k_r \right) { N ! \over \prod_{r=1}^M k_r ! } \prod_{r=1}^M
(\omega_r)^{k_r} \; e^{-J}
$$
where\cite{lipa}
$$
J = -\frac12 \sum_{r, r'} k_r \; k_{r'} \; V_{r, r'} +\frac12 \sum_r
k_r \; V_{r, r} + \frac12 \sum_{r, r', r''} k_r \; k_{r'} \; k_{r''}
\left[ < V_{r, r'} \;  V_{r, r''} > - < V_{r, r'} >\;< V_{r, r''} > \right]
+ \ldots
$$
and
$$
V_{r, r'} = \left. {1 \over \omega_r \; \omega_{r'}} {\eta \over N} 
\int_0^1 \int_0^1 { {d^3 r_1 \; d^3 r_2} \over
{ |{\vec r}_1 - {\vec r}_2|}}\right|_{{\vec r}_1 \in \, \omega_r ,\;
{\vec r}_2 \in \,\omega_{r'} }\; .
$$
Assuming $ 1/N \ll \omega_r < N^{-2/3} $ one can neglect in $ J $ terms
quadratic and higher in $  V_{r, r'} $\cite{lipa}.

The particle density is defined as
$$
N \; \rho({\vec r}) \equiv \sum_{r=1}^M{k_r \over \omega_r } \; \theta({\vec
r} \in \, \omega_r ) \; .
$$
Therefore, we can write the sums over $ r $ as integrals in the following way
$$
\frac12 \sum_{r, r'} k_r \; k_{r'} \; V_{r, r'} = {\eta \over 2 \, N} 
\int_0^1 \int_0^1 { {d^3 r_1 \; d^3 r_2} \over {|{\vec r}_1 - {\vec
r}_2|}} \; \rho({\vec r_1})\; \rho({\vec r_2}) \; .
$$
Using Stirling's' formula one finds that
$$
\prod_{r=1}^M{ (\omega_r)^{k_r}\over k_r !}\buildrel{ N \to \infty }\over = 
{1 \over N^N} \, \prod_{r=1}^M{ 1 \over \sqrt{2\pi \, k_r}} \; e^{-N
\int d^3x \; \rho({\vec x}) \; \log[\rho({\vec x})/e] }\; .
$$
Collecting all terms yields,
$$
N! \, \prod_{r=1}^M{ (\omega_r)^{k_r}\over k_r !}\; e^{-J} \buildrel{ N \to
\infty }\over =  e^{{N \; \eta \over2} \int  {d^3x  \; d^3y \over |{\vec
x}-{\vec y}|} \;\rho({\vec x}) \; \rho({\vec y}) - N  \int d^3x \; \rho({\vec
x}) \; \log[\rho({\vec x})/e]}
$$
whereas the constraint in the number of particles takes the form
$$
\delta\left(N -\sum_{r=1}^M k_r \right) = \frac1{N} \; \delta\left(\int  d^3x \,
\rho({\vec x}) - 1 \right) 
$$
and finally,
$$
e^{\Phi_N(\eta)} \buildrel{ N>>1}\over= \frac1{N} \int\int D\rho\;
e^{{N \; \eta 
\over2} \int  {d^3x  \; d^3y \over |{\vec x}-{\vec y}|} \;\rho({\vec
x}) \; \rho({\vec y}) - N  \int d^3x \; \rho({\vec x}) \;
\log[\rho({\vec x})/e]} \; \delta\left(\int  d^3x \,
\rho({\vec x}) - 1 \right) 
$$
Replacing the  Dirac delta by its Fourier representation 
$$
\frac1{N}\delta\left(\int  d^3x \,\rho({\vec x}) - 1 \right) =
\int {d{\hat a} \; \over 2 \pi} \; e^{iN{\hat a}\left(\int  d^3x
\,\rho({\vec x}) - 1 \right)} 
$$
yields eq.(\ref{zcanmf}).

\section{Calculation of the saddle point}

We prove in this Appendix that the integral 
\begin{equation}\label{Ilam}
I(\lambda) \equiv \int_0^{\lambda} x^2 \; dx \; [\chi'(x)]^2
\end{equation}
takes the value
\begin{equation}\label{solucion}
I(\lambda) = \lambda \; \eta^R \; (6 - \eta^R) - 2 \; \lambda^3  \;
e^{\chi(\lambda)} 
\end{equation}
Here $ \chi(x) $ is a regular solution of eq.(\ref{ecuaxi}) in the
interval $ 0 \leq x \leq \lambda $ fulfilling the relation
(\ref{lambaxi}). 

We start by computing the derivative of $ I(\lambda) $ in two
ways. According to the definition (\ref{Ilam})
$$
{ d I(\lambda) \over d \lambda} = \lambda^2 \;  [\chi'(\lambda)]^2
$$
Then, we compute the derivative of eq.(\ref{solucion}) with respect to $
\lambda $ and use eqs.(\ref{ecuaxi}) and (\ref{lambaxi}). We find
after calculation that both results coincide. 

Finally, we observe that both eqs.(\ref{Ilam}) and (\ref{solucion})
vanish at $ \lambda = 0 $. Therefore, eq.(\ref{solucion}) is valid.

\section{Abel's equation of first kind for the equation of state}

In the mean field approximation the equation of state for spherical
symmetry satisfies the first order differential equation (\ref{abel})
\begin{equation}\label{abelA}
\eta^R(3f_{MF}-1)f'_{MF}(\eta^R)+(3f_{MF}-3+\eta^R) f_{MF} = 0 \;.
\end{equation}
with the boundary condition $  f_{MF}(0) = 1 $.

We can solve eq.(\ref{abelA}) in power series in $ \eta^R $  around the
origin,
\begin{equation}\label{serie}
f_{MF}(\eta) = 1 + \sum_{n=1}^{\infty} f_n \; \eta^n
\end{equation}
Inserting eq.(\ref{serie}) into eq.(\ref{abelA}) yields the quadratic
recurrence relation
$$
f_n = -{1 \over 2\, n +3}\;\left[ f_{n-1} + 3 \sum_{k=2}^n k \; f_{k-1}\;
f_{n-k+1} \right] \quad \mbox{for} \; n \geq 2 \; .
$$
where $ f_1 = -\frac15 $.

We find from this recurrence relation,
$$
f_2 = - {1 \over 175} \quad , \quad f_3 = - {2 \over 1575} \quad ,
\quad f_4 = - {991 \over 3031875} 
$$
All coefficients $ f_n $ are negative rational numbers for $ n \geq
1$. They decrease very fast with $ n $ as
$$
f_n \buildrel{ n \gg 1 }\over = - {0.0956678 \ldots \over
[\eta^R_C]^n \; n^{3/2}} \left[1 + {\cal O}\left({1 \over n}\right)\right]
$$ 
This formula reproduces the large orders of the expansion of $
\sqrt{\eta^R_C - \eta^R} $ describing the behaviour of $ f_{MF}(\eta)
$ near  $ \eta^R_C $ [see eq.(\ref{cercaetac}) and ref.\cite{grad}]
$$
\sqrt{\eta^R_C - \eta^R} = -  \frac12 \sqrt{\eta^R_C \over \pi}
\sum_{n=0}^{\infty} { \Gamma(n-\frac12) \over n! } \left( {\eta^R
\over\eta^R_C } \right)^{\! n}
$$
Notice that
$$
 -  \frac12 \sqrt{\eta^R_C \over \pi}{ \Gamma(n-\frac12) \over n! }
\buildrel{ n \gg 1 }\over = -{0.447594 \ldots \over n^{3/2} } \left[1 + {\cal
O}\left({1 \over n}\right)\right] 
$$
and that $ 0.213738\ldots \times 0.447594 \ldots = 0.0956678 \ldots $.

The power series (\ref{serie}) thus has a radius of convergence $ \eta^R_C =
2.517551\ldots $. The singularity of $ f_{MF}(\eta) $ nearest to the
origin is thus the critical point.

\end{document}